\documentclass[10pt,twocolumn,letterpaper]{article}

\usepackage[pagenumbers]{cvpr}
\usepackage[T1]{fontenc}
\usepackage{amsthm}
\usepackage{mathtools}
\usepackage{enumitem}
\newtheorem{theorem}{Theorem}[section]

\newtheorem{lemma}[theorem]{Lemma}
\newtheorem{corollary}[theorem]{Corollary}
\theoremstyle{definition}

\theoremstyle{remark}

\newif\ifwithappendix
\withappendixtrue
\ifwithappendix\else\fi
\definecolor{cvprblue}{rgb}{0.21,0.49,0.74}
\usepackage[pagebackref,breaklinks,colorlinks,allcolors=cvprblue]{hyperref}
\ifwithappendix\else
  \usepackage{refcount}
  \setrefcountdefault{100000}
  \makeatletter
  \let\NNO@backcite\backcite
  \def\backcite#1#2{\NNO@filterbackcite{#1}#2}
  \def\NNO@filterbackcite#1#2#3#4{%
    \ifnum#2<\getpagerefnumber{app:theory}\relax
      \NNO@backcite{#1}{{#2}{#3}{#4}}%
    \fi}
  \makeatother
\fi

\def\confName{CVPR}
\def\confYear{2027}
\title{Discretization-Invariant Unrolled Image Reconstruction\\ with Nemytskii Neural Operators}
\author{Guixian Xu, Junqi Tang, Jinglai Li}

\begin{document}
\maketitle
\begin{abstract}
Unrolled networks alternate data-consistency updates with a learned prior
and are widely used for image reconstruction. The data-consistency update
can be re-discretized on any grid, but the learned prior is often
tied to its training grid: a network trained on one grid degrades markedly
on another, and data of different resolutions usually have to be resampled
to a common grid before joint training. 
We propose an unrolled network
formulated in function space with a Nemytskii neural operator (NNO) as the
prior. Its parameters are shared across grids, and each data-consistency
step uses the forward operator discretized on the current grid, so every
training image is used on its native grid without resampling and the same
measurements can be reconstructed on different grids. We prove that its
implementations on different grids approximate one continuum network and
that, on common training pairs, the grid-wise losses approximate a shared
continuum loss up to constants independent of the network parameters.
On sparse-view CT and accelerated MRI, NNO achieves the highest average
PSNR across grids among the priors tested. With fewer training targets at
finer resolutions, mixed-resolution training improves NNO's PSNR on the
finest grid in both tasks.
\end{abstract}

\section{Introduction}
\label{sec:intro}

Unrolled networks~\cite{monga2021unrolling,arridge2019solving} turn a fixed
number of iterations of an optimization algorithm into the stages of a
network that is trained end to end~\cite{gregor2010lista}. Each stage
combines a data-consistency update, which uses the known forward operator,
with a learned module such as  a regularizer or a proximal map.
This design has been applied to compressive sensing~\cite{zhang2018ista},
accelerated
MRI~\cite{yang2016admmnet,hammernik2018varnet,aggarwal2019modl,sriram2020e2evarnet},
CT~\cite{adler2018lpd,chen2018learn,tang2025practical}, EIT~\cite{zhou2024deep}, and image
restoration~\cite{zhang2020usrnet,mou2022dgunet}. Because the forward model
enters every stage, the learned modules can focus on prior information
about the images. Such networks are interpretable, since each stage
corresponds to one iteration, and can be trained on datasets of moderate
size~\cite{monga2021unrolling}.

In many image reconstruction problems, the unknown image is a function of
spatial position, such as the attenuation coefficient in CT, and the
forward operator (here the Radon transform) acts on such functions. The
pixel grid is a discretization chosen at reconstruction time, and the same
measurements can be reconstructed on different grids. The data-consistency
step follows from the forward model and only needs to be discretized again
on a new grid, whereas the learned prior is tied to the grid it was trained
on. Most unrolled networks, including those cited above, use convolutional
priors whose kernels have a fixed size in pixels. These networks accept
inputs of any size, but when the pixel spacing changes for a fixed field of
view, the same kernels act at a different physical scale. As a result, a
network trained on one grid performs markedly worse on
another~\cite{jatyani2025unified} (see also Section~\ref{sec:exp_single}),
and reconstructing the same measurements on a different grid requires
retraining.

A discretization-invariant unrolled network~\cite{kovachki2023neural},
whose reconstructions are consistent across grids, is of significant
practical importance. Such a network needs no retraining when the
reconstruction grid changes. A particularly important benefit of
discretization invariance is that the network can be trained on data at
several resolutions using a single grid-independent loss function. Such mixed-resolution
training is valuable in practice because imaging data often come at
different resolutions; clinical MRI varies in resolution
both within and across
centers~\cite{iglesias2023synthsr,henschel2022fastsurfervinn}.
For example, when low-resolution images are plentiful and high-resolution
ones are few, as in MRI, where high-resolution scans take longer,
mixed-resolution training can use both instead of only the few
high-resolution examples.

One can enable mixed-resolution training by resampling all training
images to a common grid beforehand. This strategy is commonly used in medical
image analysis~\cite{isensee2021nnunet,henschel2022fastsurfervinn}.
Resampling, however, does not remove the grid dependence of the prior, and
upsampling may introduce artifacts while downsampling discards
detail~\cite{henschel2022fastsurfervinn}.

A mathematically more rigorous treatment is  to formulate the unrolled network in the
continuum, with the prior represented by a neural
operator~\cite{kovachki2023neural}; each grid then provides only a
discretization of the same network.
Neural operators are designed to learn maps between function spaces, so
that their discretizations on different grids approximate the same
continuum operator~\cite{kovachki2023neural}. They were originally introduced for learning solution operators of
PDEs~\cite{lu2021deeponet,li2021fno} and have since been applied to vision
tasks such as image classification~\cite{kabri2023resolution},
segmentation~\cite{guibas2022afno,liu2024resolution}, and
super-resolution~\cite{wei2023srno}.
In inverse imaging, Auras et al.~\cite{auras2026unets} view U-Nets as neural
operators and study how post-processing networks for limited-angle CT
generalize across resolutions, including training on mixed resolutions. Unrolled networks with neural operators have been proposed for
MRI and sparse-view CT: Jatyani et al.~\cite{jatyani2025unified} use neural
operators in both the image and measurement spaces, so that one model
handles different undersampling patterns, and CTO~\cite{datta2025cto} uses
them in the sinogram and image spaces, so that one model handles different
sampling rates. Both methods are trained on a single grid; when tested on a finer grid, they bilinearly
interpolate an intermediate image to that grid and are evaluated against
bilinearly interpolated ground truth.
The discretization invariance considered here should not be confused with
the resolution agnosticism of these works. Resolution agnosticism means
that one model handles different sampling patterns, sampling rates, or image
sizes; discretization invariance means that reconstructions of the same
measurements on different grids approximate the same continuum image. To our
knowledge, the mixed-resolution training that uses each
training image directly on its native grid has not been applied to
unrolled networks. For these networks, theory on resolution changes covers individual
layers~\cite{auras2026unets}, but not a complete unrolled network with its
data-consistency steps, or its training losses on different grids.

We propose an unrolled network formulated in function space
with a Nemytskii neural operator (NNO) as the prior. The NNO composes
pointwise maps with the Fourier transform and its inverse; its spectral maps
take the physical frequency as input and act only on the frequency band seen
in training. The same parameters are used on different grids, and each
data-consistency step uses the forward operator discretized on the current
grid, so every training image can be used on its own grid and the same
measurements can be reconstructed on different grids
(Section~\ref{sec:method}). We prove that the
implementations of the network on different grids approximate one continuum
network, and that on common training pairs the grid-wise losses approximate
a shared continuum loss up to constants independent of the network parameters
(Section~\ref{sec:theory}). In experiments on
sparse-view CT and accelerated MRI, NNO trained on a coarse grid is the most
accurate prior on a grid four times finer, and its average accuracy across
grids is the best among the priors in both unrolled schemes we test.
With fewer supervision images at finer resolutions, mixed-resolution
training improves NNO's PSNR on the finest grid in both tasks. NNO achieves
the highest PSNR there and on the unseen intermediate grid
(Section~\ref{sec:experiments}).

\section{Method}
\label{sec:method}

\subsection{Function-Space Unrolling for Image Reconstruction}
\label{sec:function_space_unrolling}

We formulate the image reconstruction process in a function space.
Let $\Omega \subset \mathbb{R}^{d}$ be a bounded physical imaging
domain, and represent the unknown image as a real-valued function
$u^\dagger \in \mathcal{X}=L^{2}(\Omega;\mathbb{R})$.
The imaging process is modeled by a bounded linear forward operator
$\mathcal{A}:\mathcal{X}\to\mathcal{Y}$, where
$\mathcal{Y}=\mathbb{R}^{m}$ is the measurement space
and $m$ is the number of measurements. Specifically, we have
\begin{equation}
    y=\mathcal{A}u^\dagger+\eta,
    \label{eq:reconstruction_measurements}
\end{equation}
where $\eta\in\mathcal{Y}$ denotes measurement noise.
Image reconstruction seeks to estimate $u^\dagger$ from $y$.
The image is commonly reconstructed by solving the following regularized optimization problem:
\begin{equation}
\min_{u\in\mathcal X}
\frac12\|\mathcal A u-y\|^{2}
+\lambda\,\mathcal R(u), \label{eq:regopt}
\end{equation}
where \(\mathcal R\) is an image regularizer and \(\lambda>0\) controls the regularization strength.

In practice, selecting an appropriate regularizer $\mathcal{R}$ and
tuning the regularization parameter $\lambda$ are often challenging
and require problem-specific expertise.
This motivates unrolled reconstruction, which unfolds a fixed number of optimization iterations into a sequence of trainable mappings.
Through joint training of these mappings, the resulting model learns an implicit image prior from data.

Conventional unrolled reconstruction methods typically operate on
finite-dimensional image representations obtained through discretization.
Here, we adopt a formulation defined directly on the image function
space $\mathcal{X}$.
The unrolled iterations take the form
\begin{equation}
\begin{aligned}
    v^k &= \mathcal{G}_k(u^k;y,\mathcal{A}), \\
    u^{k+1} &= \mathcal{D}_{\theta_k}(v^k),
\end{aligned}
\qquad k=0,\ldots,K-1,
\label{eq:unrolled_iterations}
\end{equation}
where $u^0$ is an initial reconstruction,
$\mathcal{G}_k$ denotes a data-consistency update based on the
forward operator $\mathcal{A}$, and $\mathcal{D}_{\theta_k}$
is the prior update.
Both $\mathcal{G}_k(\cdot;y,\mathcal{A})$
and $\mathcal{D}_{\theta_k}$ are operators
from $\mathcal{X}$ to $\mathcal{X}$, with $\mathcal{G}_k$
typically fixed by the physical model and optimization algorithm
and $\mathcal{D}_{\theta_k}$ learned from data.
We therefore parameterize $\mathcal{D}_{\theta_k}$ as a neural
operator to realize the learned prior update in function space.

\subsection{Nemytskii Neural Operator}
\label{sec:nno}
\begin{figure*}[t]
\centering
\includegraphics[width=\textwidth]{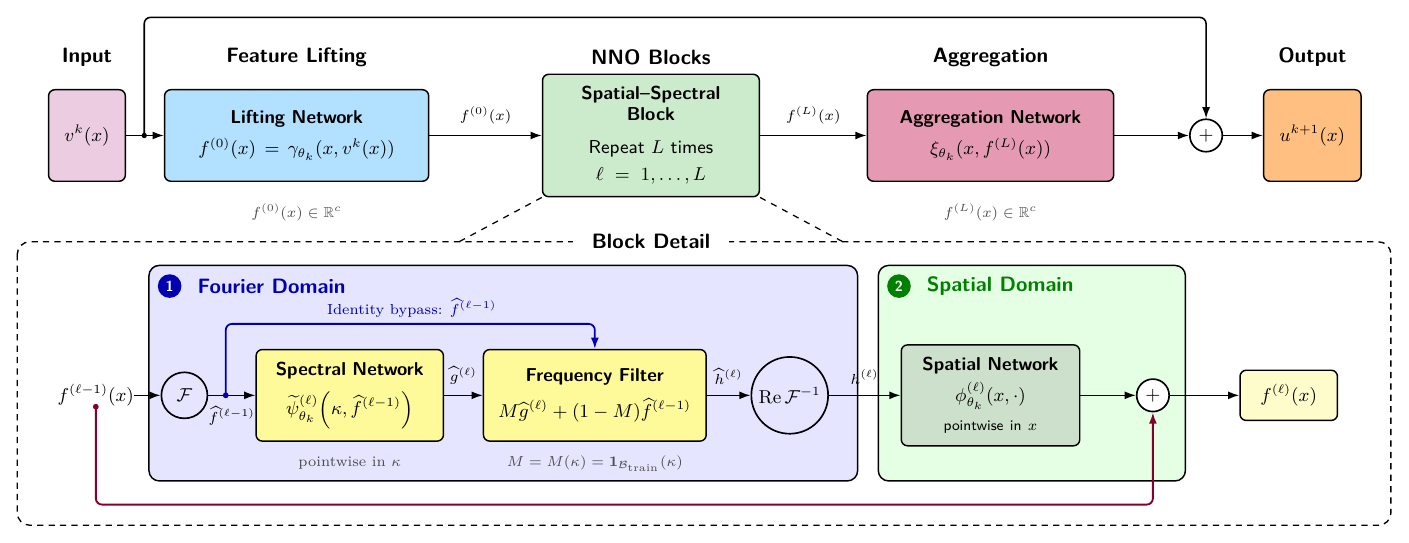}
\caption{NNO prior in stage $k$ of the unrolled network,
$u^{k+1}=\mathcal D_{\theta_k}(v^k)$. Top: lifting, $L$ spatial--spectral
blocks and aggregation, with a residual connection. Bottom: one
spatial--spectral block~\eqref{eq:nno_block}; the frequency filter leaves the
coefficients outside $\mathcal B_{\mathrm{train}}$ unchanged.}
\label{fig:nno}
\end{figure*}

In principle, any neural operator mapping $\mathcal{X}$ to
$\mathcal{X}$ can be incorporated into this unrolling scheme.
Here, we propose a Nemytskii neural operator (NNO) built by
composing learnable pointwise mappings with the Fourier
transform and its inverse (Figure~\ref{fig:nno}).

A Nemytskii operator acts on a function by applying a
finite-dimensional mapping pointwise. Let
$n:\Omega\times\mathbb{R}^{c_{\mathrm{in}}}
\to\mathbb{R}^{c_{\mathrm{out}}}$
be a mapping. The associated Nemytskii operator $\mathcal{N}$
acts on a field $f:\Omega\to\mathbb{R}^{c_{\mathrm{in}}}$
pointwise according to
\begin{equation}
    (\mathcal{N}f)(x)=n(x,f(x)),
    \label{eq:nemytskii_operator}
\end{equation}
yielding an output field
$\mathcal{N}f:\Omega\to\mathbb{R}^{c_{\mathrm{out}}}$.
The dependence on $x$ allows position-dependent mappings.
The same construction applies in the Fourier domain, where
the mapping acts pointwise on Fourier coefficients at each
frequency $\kappa$.
Complex-valued features are represented by their real and
imaginary parts.

We use these Nemytskii operators as building blocks and compose
them with Fourier transforms to capture global spatial
dependencies.
Given an input $v\in\mathcal{X}$,
a lifting operator first maps it to a feature field with $c$
channels:
\begin{equation}
    f^{(0)}(x)
    = \gamma_{\theta}(x,v(x)),
    \label{eq:nno_lifting}
\end{equation}
where $\gamma_{\theta}:\Omega\times\mathbb{R}\to\mathbb{R}^{c}$
is a neural network parameterized by $\theta$, and $c$ denotes
the number of feature channels.
The lifted features are then processed by $L$ spatial--spectral
blocks. For $\ell=1,\ldots,L$, each block takes the form
\begin{subequations}
\label{eq:nno_block}
\begin{align}
    \widehat{f}^{(\ell-1)}
    &= \mathcal{F}f^{(\ell-1)}, \label{eq:nno_block_a} \\
    \widehat{h}^{(\ell)}(\kappa)
    &= \psi_{\theta}^{(\ell)}\bigl(\kappa,\widehat{f}^{(\ell-1)}(\kappa)\bigr), \label{eq:nno_block_b} \\
    h^{(\ell)}
    &= \operatorname{Re}\,\mathcal{F}^{-1}\widehat{h}^{(\ell)}, \label{eq:nno_block_c} \\
    f^{(\ell)}(x)
    &= f^{(\ell-1)}(x)+\phi_{\theta}^{(\ell)}\bigl(x,h^{(\ell)}(x)\bigr), \label{eq:nno_block_d}
\end{align}
\end{subequations}
where $\mathcal{F}$ and $\mathcal{F}^{-1}$ denote the Fourier transform
and its inverse, and $\kappa$ denotes a frequency coordinate.
The spectral network
$\psi_{\theta}^{(\ell)}:\mathbb{R}^{d}\times\mathbb{C}^{c}\to\mathbb{C}^{c}$
transforms the feature channels at each frequency, and the spatial network
$\phi_{\theta}^{(\ell)}$ performs pointwise refinement in the spatial
domain through a residual update.
A pointwise aggregation network then maps the terminal features to a
real-valued residual correction,
\begin{equation}
    \mathcal{D}_{\theta}(v)(x)=v(x)+\xi_{\theta}\bigl(x,f^{(L)}(x)\bigr),
    \label{eq:nno_prior}
\end{equation}
where $\xi_{\theta}:\Omega\times\mathbb{R}^{c}\to\mathbb{R}$.

For the spectral network $\psi_{\theta}^{(\ell)}$, we further incorporate a
frequency filter that restricts the learned map to the frequencies observed
during training:
\begin{equation}
    \psi_{\theta}^{(\ell)}(\kappa,z)
    ={} M(\kappa)\,\widetilde{\psi}_{\theta}^{(\ell)}(\kappa,z) 
    +\bigl(1-M(\kappa)\bigr)\,z .
\end{equation}
Here $\widetilde{\psi}_{\theta}^{(\ell)}$ is the learned network,
$M=\mathbf{1}_{\mathcal{B}_{\mathrm{train}}}$ is the indicator function of
$\mathcal{B}_{\mathrm{train}}=\{\kappa:\|\kappa\|_{\infty}<\kappa_{\mathrm{train}}^{\max}\}$,
which is fixed in physical frequency and shared by all channels.
Coefficients outside $\mathcal{B}_{\mathrm{train}}$ therefore pass through
unchanged, and the network does not act on frequency coordinates unseen in
training.

Finally we insert NNO into the unrolling formulation
and  use it as the prior update in \eqref{eq:unrolled_iterations},
\begin{equation}
    u^{k+1}=\mathcal{D}_{\theta_k}(v^k),
    \qquad k=0,\ldots,K-1,
    \label{eq:nno_unrolling}
\end{equation}
where $\theta_k$ denotes the NNO parameters at stage $k$ and
$v^k=\mathcal{G}_k(u^k;y,\mathcal{A})$ is the data-consistency update.

\subsection{Discretization and Training}
\label{sec:disc_train}

We discuss the training process in a supervised manner. 
Let $u^{K}(y;\Theta)$ denote the output of \eqref{eq:unrolled_iterations},
where $\Theta$ collects $\{\theta_k\}_{k=0}^{K-1}$ and any learnable
parameters of $\mathcal{G}_k$.
Given training images $u_1^\dagger,\dots,u_n^\dagger\in\mathcal{X}$ and noise
realizations $\eta_1,\dots,\eta_n\in\mathcal{Y}$, we form the training pairs
$\{(y_i,u_i^\dagger)\}_{i=1}^{n}$ with $y_i=\mathcal{A}u_i^\dagger+\eta_i$.
All stages are trained jointly by minimizing the empirical risk
\begin{equation}
    \min_{\Theta}\; \mathcal{L}(\Theta):=\frac{1}{n}\sum_{i=1}^{n}
    \bigl\|u^{K}(y_i;\Theta)-u_i^\dagger\bigr\|_{\mathcal{X}}^{2},
    \label{eq:training_loss}
\end{equation}
with backpropagation through all $K$ stages.

In practice, the images $u_i^\dagger$ in the training set are only available
as finite-resolution arrays, so training can only be performed on such
discrete data.
Let $\Gamma_N\subset\Omega$ be a uniform grid of $N^{d}$ points with cell
volume $h_N^{d}$, let $\mathcal{S}_N$ sample a function on $\Gamma_N$, and
let $T_N$ remove the Fourier modes that $\Gamma_N$ cannot represent.
For a fixed resolution $N$, the training pairs become
$\{(y_i,u_{i,N}^\dagger)\}_{i=1}^{n}$, where
$u_{i,N}^\dagger=\mathcal{S}_NT_Nu_i^\dagger\in\mathbb{R}^{N^{d}}$ is the
band-limited representation of $u_i^\dagger\in\mathcal{X}$ on $\Gamma_N$,
approximated in practice by anti-aliased resampling of the
highest-resolution image.
Let $\mathcal{A}_N:\mathbb{R}^{N^{d}}\to\mathbb{R}^{m}$ be the discretization
of $\mathcal{A}$ on $\Gamma_N$, and let $\mathcal{A}_N^{*}$ be its adjoint
with respect to the inner product
$\langle a,b\rangle_N=h_N^{d}\sum_{j}a_jb_j$, which induces the norm
$\|\cdot\|_N$.
Let $u_N^{K}(y;\Theta)$ denote the output of \eqref{eq:unrolled_iterations}
on $\Gamma_N$, where $\mathcal{A}$, $\mathcal{A}^{*}$ and
$\mathcal{D}_{\theta_k}$ are replaced by $\mathcal{A}_N$, $\mathcal{A}_N^{*}$
and their grid versions with the same parameters $\Theta$.
Given the training pairs $\{(y_i,u_{i,N}^\dagger)\}_{i=1}^{n}$, the
$N$-resolution approximation of $\mathcal{L}(\Theta)$ is
\begin{equation}
    \mathcal{L}_N(\Theta)
    =\frac{1}{n}\sum_{i=1}^{n}
    \bigl\|u_N^{K}(y_i;\Theta)-u_{i,N}^\dagger\bigr\|_{N}^{2}.
    \label{eq:training_loss_N}
\end{equation}
If training pairs are available at several resolutions $N\in\mathcal{T}$,
they can be combined into a single mixed-resolution objective, since the
parameters $\Theta$ do not depend on $N$:
\begin{equation}
    \mathcal{L}_{\mathrm{mix}}(\Theta)
    =\sum_{N\in\mathcal{T}}\pi_N\,\mathcal{L}_N(\Theta),
    \qquad \pi_N\ge0,\ \ \sum_{N\in\mathcal{T}}\pi_N=1.
    \label{eq:training_loss_mix}
\end{equation}
In practice, each mini-batch contains samples from several resolutions, and
the losses of the resolutions are weighted by their numbers of samples, so
$\pi_N$ is the expected fraction of resolution-$N$ samples in a mini-batch.

\section{Theoretical Analysis}
\label{sec:theory}
A key theoretical question is whether the finite-resolution implementation of
the proposed network approximates its continuum formulation. 
We answer it at two levels: the
reconstruction operator (Section~\ref{sec:discretization_consistency}) and
the training loss (Section~\ref{sec:mixed_training}). Proofs are in
Appendix~\ref{app:theory}.

\subsection{Discretization Consistency}
\label{sec:discretization_consistency}

We identify $\Omega$ with the torus $\mathbb T^d$ and write
$\widehat w(\kappa)$, $\kappa\in\mathbb Z^d$, for Fourier coefficients. On
$\Gamma_N$ let
$a_N[v](\kappa)=h_N^d\sum_jv_je^{-2\pi i\kappa\cdot x_j}$ be the
forward-normalized DFT, with $\kappa$ in the set $K_N$ of frequencies
representable on $\Gamma_N$, and
$a_{N,F}(w)=\|a_N[\mathcal S_Nw]-\widehat w\,\|_{\ell^2(F)}$ the energy that
sampling folds into a finite set $F$. The measurement $y$
in~\eqref{eq:reconstruction_measurements} is shared by all grids. The
initialization $u^0$ is a fixed linear reconstruction from $y$ (zero
filling in MRI, FBP in CT) and the grid uses its samples
$\mathcal S_Nu^0$; $\mathcal G_{k,N}$ denotes the grid version of
$\mathcal G_k$.

\begin{theorem}
\label{thm:local}
Assume that the grid updates $\mathcal G_{k,N}$ are Lipschitz uniformly
in $N$,  that $\gamma_{\theta_k}$,
$\phi^{(\ell)}_{\theta_k}$, $\xi_{\theta_k}$ are smooth with bounded
derivatives and $\psi^{(\ell)}_{\theta_k}(\kappa,\cdot)$ is Lipschitz, and
that $\mathcal B_{\mathrm{train}}\subset K_N$ for every grid considered.
\begin{enumerate}
\item[(a)] For every $N$,
\begin{equation}
\begin{aligned}
    &\big\|u^K_N(y;\Theta)-\mathcal S_Nu^K(y;\Theta)\big\|_N\\
    &\le\sum_{k=0}^{K-1}\Lambda'_k\,\varepsilon^{\mathrm{data}}_k(N)
    +\sum_{w\in W}\Lambda_w\,a_{N,\mathcal B_{\mathrm{train}}}(w),
\end{aligned}
\label{eq:th_localization}
\end{equation}
where $W$ collects the continuum feature fields $f^{(\ell-1)}$ entering
the Fourier transform~\eqref{eq:nno_block_a} in all blocks and stages,
  $ \varepsilon^{\mathrm{data}}_k(N)
    =\|\mathcal G_{k,N}(\mathcal S_Nu^k)-\mathcal S_N\mathcal G_k(u^k)\|_N$ 
is the consistency error of the data-consistency update at the continuum
iterate $u^k$, and $\Lambda_w,\Lambda'_k$ are Lipschitz propagation
factors independent of $N$.
\item[(b)] If the continuum fields are $C^\infty(\mathbb T^d)$, every
spectral term in~\eqref{eq:th_localization} is $O(N^{-\beta})$ for every
$\beta>0$; if they are piecewise Lipschitz with finitely many $C^1$
interfaces, every spectral term is $O(N^{-1})$.
\end{enumerate}
\end{theorem}
\paragraph{Interpretation.}
Theorem~\ref{thm:local} (a) shows that the proposed network defines one continuum
operator whose implementations on different grids differ only through two
errors. The first comes from discretizing the forward operator in the
data-consistency updates and is present for any prior.
The second comes from the prior: detail that its pointwise networks
create beyond the resolution of the grid is partly aliased into the low frequencies
processed by the spectral networks.
By part~(b), as the grid is refined,  the aliasing error decreases at a rate determined by the regularity of the intermediate features.

A special case is Cartesian MRI.
Here, the frequency mixing behind the second error also enables
the network to estimate unmeasured k-space coefficients from the data:
\begin{corollary}
\label{cor:exact}
For Cartesian MRI, $\mathcal A=\mathbf 1_{\mathcal P}\mathcal F$,
discretized as $\mathcal A_N=\mathbf 1_{\mathcal P}a_N$ with
$\pm\mathcal P\subset K_N$, and gradient or proximal data-consistency
updates: if $\gamma_{\theta_k}$, $\phi^{(\ell)}_{\theta_k}$ and
$\xi_{\theta_k}$ are affine and position-independent, then
$u^K_N(y;\Theta)=\mathcal S_Nu^K(y;\Theta)$, and the output coefficients
outside $\pm\mathcal P$ are independent of $y$.
\end{corollary}

\subsection{Mixed-Resolution Training}
\label{sec:mixed_training}

  With $I_N$ the trigonometric interpolation on $\Gamma_N$, let
$\widehat\varepsilon_N^2=\frac1n\sum_i\|(I-T_N)u_i^\dagger\|^2$ be the
sub-pixel energy of the targets, which no output on $\Gamma_N$ can fit, and
$\delta_N^2=\frac1n\sum_i\|I_Nu^K_N(y_i;\Theta)-u^K(y_i;\Theta)\|^2$ the
distance between the interpolated grid outputs and the continuum outputs;
$\delta_N$ is at most the root-mean-square node-wise error of
Theorem~\ref{thm:local} plus the representability error
$\big(\frac1n\sum_i\|(I-I_N\mathcal S_N)u^K(y_i;\Theta)\|^2\big)^{1/2}$.

\begin{theorem}
\label{thm:consistency}
Under the assumptions of Theorem~\ref{thm:local}, for every $\Theta$ and
every grid size $N$:
\begin{enumerate}
   \item[(a)]
   \begin{align*}
       &\big|\mathcal L_N+\widehat\varepsilon_N^2-\mathcal L\big|
       \le\delta_N\big(2\sqrt{\mathcal L}+\delta_N\big),\\
       &\Big|\mathcal L_{\mathrm{mix}}
       +\textstyle\sum_{N\in\mathcal T}\pi_N\widehat\varepsilon_N^2
       -\mathcal L\Big|\\
       &\qquad\le\textstyle\sum_{N\in\mathcal T}\pi_N\,
       \delta_N\big(2\sqrt{\mathcal L}+\delta_N\big).
   \end{align*}
\item[(b)] For all $N,N'$, including grids not used in training,
\[
    \Big|\sqrt{\mathcal L_N+\widehat\varepsilon_N^2}
    -\sqrt{\mathcal L_{N'}+\widehat\varepsilon_{N'}^2}\Big|
    \le\delta_N+\delta_{N'}.
\]
\end{enumerate}
\end{theorem}

\paragraph{Interpretation.}
This theorem addresses whether minimizing the resolution-$N$ loss
$\mathcal L_N$ or the mixed-resolution loss $\mathcal L_{\mathrm{mix}}$
approximately minimizes their continuum counterpart $\mathcal L$
in~\eqref{eq:training_loss}.
Part~(a) states that, up to constants independent of $\Theta$,
$\mathcal L_N$ and $\mathcal L_{\mathrm{mix}}$ differ from $\mathcal L$ by
an error controlled by $\delta_N$; when $\delta_N$ is small uniformly in
$\Theta$, minimizing them therefore approximately minimizes $\mathcal L$.
Part~(b) implies that, up to a constant, the losses at different
resolutions differ only by small discretization errors, so parameters
that fit well at one resolution also fit well at the others, including
resolutions not used in training.

\section{Experiments}
\label{sec:experiments}

We evaluate the proposed network on sparse-view CT and accelerated MRI under two
training protocols: single-resolution training
(Section~\ref{sec:exp_single}), where each model is trained on one grid and
evaluated on it and on finer grids, the setting of Theorems~\ref{thm:local}
and~\ref{thm:consistency}(b); and mixed-resolution training with
$\mathcal L_{\mathrm{mix}}$ (Section~\ref{sec:exp_mixed}), with fewer
training targets available at finer resolutions.

\subsection{Datasets and Setup}
\label{sec:exp_setup}

\paragraph{Datasets.}
For CT, we use 120 slices of the SIIM CT
dataset~\cite{mader2016ctmedicalimages} (from
TCGA-LUAD~\cite{albertina2016tcgaluad,clark2013tcia}). For MRI, we use one
slice from each of 120 SKM-TEA knee scans~\cite{desai2021skm}, with the two
qDESS echoes combined by root-sum-of-squares. Both datasets have native
resolution $512^2$ and use the grid sizes $N\in\{128,256,384,512\}$; each
is split into 80 training and 40 test slices with no patient in common, and
each test slice comes from a different patient.
Targets on each grid are formed as in Section~\ref{sec:disc_train}.

\paragraph{Measurements.}
Measurements are simulated from the $512^2$ images: for CT, we keep 36 of
720 parallel-beam views; for MRI, single-coil k-space is undersampled by a
fixed 1D Cartesian mask with $R=4$. No noise is added. All grids use the
same $y$.

\paragraph{Schemes and baselines.}
Unless stated otherwise, we use ISTA-Net+~\cite{zhang2018ista}, whose
data-consistency update $\mathcal G_k$ is a gradient step;
Section~\ref{sec:exp_single} also uses MoDL~\cite{aggarwal2019modl}, whose
$\mathcal G_k$ is a proximal step computed by conjugate gradients. We
compare NNO with a CNN and the neural operators FNO~\cite{li2021fno},
AFNO~\cite{guibas2022afno}, and CNO~\cite{raonic2023cno}. Within each
scheme, only $\mathcal D_{\theta_k}$ differs between methods, and all priors
have parameter counts comparable to or larger than NNO's and the same
training budget.
For CT, we also compare with CTO~\cite{datta2025cto}, a resolution-agnostic
CT reconstruction method, parameter-matched and at its original size
(CTO-full, 172.5M parameters). For MRI, we compare with the unrolled
neural-operator network of Jatyani et al.~\cite{jatyani2025unified}, denoted
NO-VarNet, parameter-matched and at its original size (NO-VarNet-full,
139.7M parameters). FBP and zero-filled reconstruction serve as
learning-free references. Architectures and training details are in
Appendix~\ref{app:exp_details}.

\subsection{Single-Resolution Training}
\label{sec:exp_single}

\begin{figure*}[tp]
\centering
\begin{subfigure}{\textwidth}
  \centering
  \includegraphics[width=0.88\linewidth]{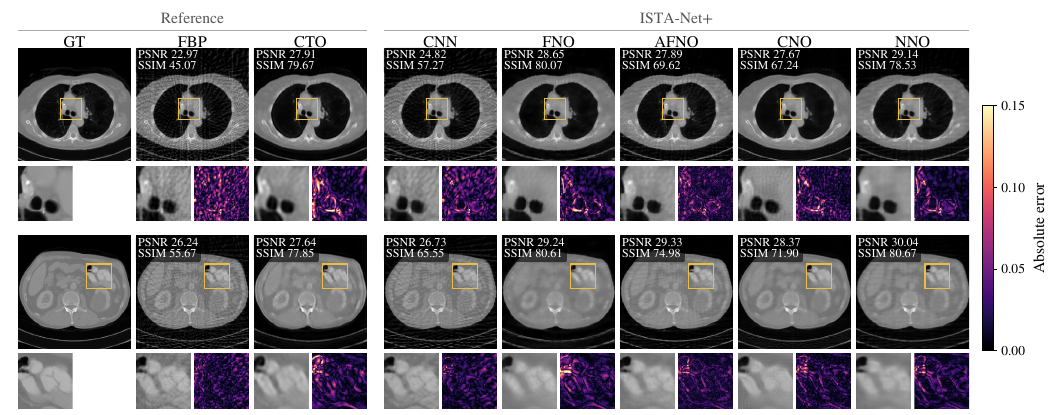}
  \caption{Sparse-view CT}
  \label{fig:single_quality_ct}
\end{subfigure}

\vspace{6pt}
\begin{subfigure}{\textwidth}
  \centering
  \includegraphics[width=0.88\linewidth]{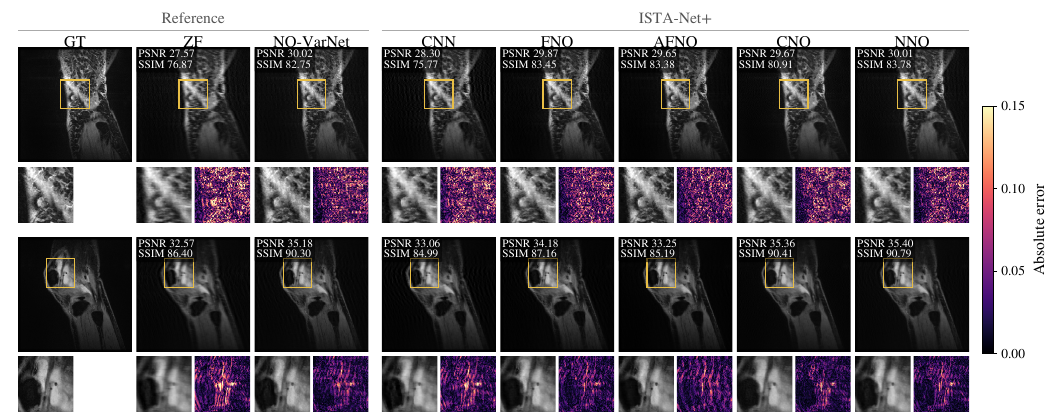}
  \caption{Accelerated MRI}
  \label{fig:single_quality_mri}
\end{subfigure}
\caption{Reconstructions with ISTA-Net+ under single-resolution training on
$128^2$, evaluated on $512^2$. CTO and NO-VarNet denote the parameter-matched versions.
Each image shows its PSNR (dB) and SSIM ($\times100$); the insets show a
zoomed region and its absolute error.}
\label{fig:single_quality}
\end{figure*}

Each model is trained on $N=128$ and evaluated on it and on the finer grids
$N=256$ and $N=512$ (Tables~\ref{tab:ct_single} and~\ref{tab:mri_single}).
On both tasks, NNO has the best average PSNR among the priors and the
best PSNR on $512^2$, four times the training resolution. It is also the
only prior whose PSNR is within 0.8~dB of the best prior on every grid.
Figure~\ref{fig:single_quality} shows examples on $512^2$, where NNO has the
highest PSNR among the priors in all four cases.

\paragraph{Sparse-view CT.}
NNO also has the best average PSNR of all methods, including CTO-full. The
methods that are most accurate on the training grid fall behind NNO on the
finer grids: CTO-full and CNO lead NNO by 1.04 and 0.18~dB on $128^2$, but
trail it by 1.17 and 1.54~dB on $512^2$. CTO-full, with about $545\times$
more parameters, retains higher SSIM. The CNN prior
loses 8.03~dB from $128^2$ to $512^2$, since its kernels are defined in
pixels and shrink in physical extent on finer grids; on $512^2$, NNO exceeds
it by 3.61~dB. FNO is stable across grids but at a low level: NNO is
4.45~dB more accurate on $128^2$ and still 0.65~dB more accurate on
$512^2$.

\paragraph{Accelerated MRI.}
NNO has the best average PSNR and the best results on $512^2$ in both
metrics; AFNO has a higher average SSIM (89.57 vs.\ 89.31). As on CT, the
priors that lead on the coarser grids fall behind on $512^2$: CNO leads on
$128^2$ by 0.19~dB and AFNO on $256^2$ by 0.72~dB, but they trail NNO by
0.10 and 0.93~dB on $512^2$. The CNN prior
again fails to transfer: on $512^2$, its SSIM is below that of zero filling.

\paragraph{Cross-grid consistency.}
To compare reconstructions of the same measurements on different grids
without ground truth, we downsample the $512^2$ reconstruction to $128^2$
by Lanczos interpolation and compute its PSNR to the $128^2$ reconstruction
(Table~\ref{tab:consistency}). NNO is the most consistent prior on CT and
the second on MRI, after CNO, which computes on a fixed $128^2$ grid
(Appendix~\ref{app:priors}). FNO is also consistent but less accurate
(Tables~\ref{tab:ct_single} and~\ref{tab:mri_single}), and the CNN prior is
the least consistent on both tasks.

\paragraph{Other unrolling schemes.}
Table~\ref{tab:schemes} compares the grid-averaged results within ISTA-Net+
and MoDL; per-grid MoDL results are in Appendix~\ref{app:exp_results}. NNO
is the only prior that ranks first in average PSNR in all four settings, and
it has the best average SSIM in two of them; its gain over the CNN prior
ranges from 0.47 to 1.85~dB. The second-best prior changes with the task
and the scheme: AFNO in three settings and CNO on MRI within MoDL. The
advantage of NNO is therefore a property of the prior and holds for both
types of data-consistency update.

\begin{table*}[t]
\centering
\caption{Sparse-view CT with ISTA-Net+ under single-resolution training on
$128^2$. Entries are PSNR (dB) / SSIM ($\times100$), mean $\pm$ std over
the test set; the average is over the three grids. Best in \textbf{bold},
second best \underline{underlined}.}
\label{tab:ct_single}
\small
\setlength{\tabcolsep}{7pt}
\resizebox{\textwidth}{!}{%
\begin{tabular}{lcccc}
\toprule
Method & $128^2$ (training grid) & $256^2$ & $512^2$ & Average \\
\midrule
FBP
& $22.66 \pm 1.86$ / $48.09 \pm 6.26$
& $23.09 \pm 1.97$ / $41.13 \pm 6.03$
& $22.98 \pm 2.05$ / $40.80 \pm 7.40$
& $22.91$ / $43.34$ \\
CNN
& $32.18 \pm 1.95$ / $90.29 \pm 2.33$
& $28.02 \pm 2.13$ / $72.33 \pm 5.70$
& $24.15 \pm 1.98$ / $53.74 \pm 7.61$
& $28.12$ / $72.12$ \\
\midrule
FNO
& $28.79 \pm 1.69$ / $81.92 \pm 2.65$
& $28.03 \pm 1.66$ / $76.54 \pm 4.13$
& $\underline{27.11 \pm 1.81}$ / $72.22 \pm 8.58$
& $27.98$ / $76.89$ \\
AFNO
& $31.46 \pm 1.81$ / $88.59 \pm 2.48$
& $\mathbf{29.10 \pm 1.79}$ / $79.89 \pm 5.00$
& $26.85 \pm 1.88$ / $64.37 \pm 7.48$
& $29.14$ / $77.62$ \\
CNO
& $\underline{33.42 \pm 2.04}$ / $\underline{92.43 \pm 2.23}$
& $27.23 \pm 1.90$ / $60.05 \pm 9.52$
& $26.22 \pm 1.87$ / $55.28 \pm 11.54$
& $28.96$ / $69.25$ \\
NNO
& $33.24 \pm 2.01$ / $91.02 \pm 2.09$
& $\underline{28.92 \pm 1.69}$ / $\underline{80.92 \pm 4.26}$
& $\mathbf{27.76 \pm 1.93}$ / $71.58 \pm 8.51$
& $\mathbf{29.97}$ / $\underline{81.17}$ \\
\midrule
CTO
& $32.35 \pm 1.77$ / $90.85 \pm 2.41$
& $27.80 \pm 1.49$ / $79.80 \pm 4.72$
& $26.29 \pm 1.59$ / $\underline{72.85 \pm 8.88}$
& $28.81$ / $\underline{81.17}$ \\
CTO-full
& $\mathbf{34.28 \pm 2.05}$ / $\mathbf{93.05 \pm 2.14}$
& $28.31 \pm 1.63$ / $\mathbf{82.45 \pm 4.67}$
& $26.59 \pm 1.73$ / $\mathbf{75.33 \pm 9.32}$
& $\underline{29.73}$ / $\mathbf{83.61}$ \\
\bottomrule
\end{tabular}}
\end{table*}

\begin{table*}[t]
\centering
\caption{Accelerated MRI with ISTA-Net+ under single-resolution training on
$128^2$. Entries are PSNR (dB) / SSIM ($\times100$), mean $\pm$ std over
the test set; the average is over the three grids. Best in \textbf{bold},
second best \underline{underlined}.}
\label{tab:mri_single}
\small
\setlength{\tabcolsep}{7pt}
\resizebox{\textwidth}{!}{%
\begin{tabular}{lcccc}
\toprule
Method & $128^2$ (training grid) & $256^2$ & $512^2$ & Average \\
\midrule
Zero-filled
& $30.89 \pm 2.72$ / $87.24 \pm 3.78$
& $29.46 \pm 2.75$ / $80.49 \pm 5.67$
& $28.85 \pm 2.78$ / $76.32 \pm 6.88$
& $29.73$ / $81.35$ \\
CNN
& $36.14 \pm 3.24$ / $95.67 \pm 2.84$
& $31.50 \pm 2.78$ / $85.31 \pm 5.59$
& $29.40 \pm 2.68$ / $75.34 \pm 6.34$
& $32.35$ / $85.44$ \\
\midrule
FNO
& $34.57 \pm 3.08$ / $93.15 \pm 2.91$
& $32.82 \pm 3.06$ / $87.99 \pm 4.82$
& $31.22 \pm 2.79$ / $\underline{82.48 \pm 5.29}$
& $32.87$ / $87.87$ \\
AFNO
& $36.36 \pm 3.17$ / $95.96 \pm 1.81$
& $\mathbf{33.96 \pm 3.06}$ / $\mathbf{90.82 \pm 3.83}$
& $30.69 \pm 2.41$ / $81.93 \pm 4.18$
& $\underline{33.67}$ / $\mathbf{89.57}$ \\
CNO
& $\underline{36.59 \pm 3.30}$ / $\underline{96.37 \pm 1.97}$
& $32.78 \pm 3.07$ / $87.06 \pm 5.21$
& $\underline{31.52 \pm 3.00}$ / $81.99 \pm 6.47$
& $33.63$ / $88.47$ \\
NNO
& $36.40 \pm 3.29$ / $95.63 \pm 2.27$
& $33.24 \pm 3.18$ / $89.07 \pm 4.89$
& $\mathbf{31.62 \pm 3.06}$ / $\mathbf{83.24 \pm 6.26}$
& $\mathbf{33.75}$ / $\underline{89.31}$ \\
\midrule
NO-VarNet
& $34.77 \pm 2.94$ / $94.46 \pm 2.07$
& $32.63 \pm 2.90$ / $87.62 \pm 4.72$
& $31.15 \pm 2.90$ / $80.97 \pm 6.55$
& $32.85$ / $87.68$ \\
NO-VarNet-full
& $\mathbf{37.09 \pm 3.30}$ / $\mathbf{96.59 \pm 1.62}$
& $\underline{33.57 \pm 3.10}$ / $\underline{90.51 \pm 3.19}$
& $28.00 \pm 2.98$ / $69.36 \pm 4.53$
& $32.89$ / $85.49$ \\
\bottomrule
\end{tabular}}
\end{table*}

\begin{table}[t]
\centering
\caption{Cross-grid consistency under single-resolution training on
$128^2$ with ISTA-Net+: PSNR (dB) between the reconstructions on $128^2$
and on $512^2$, the latter downsampled to $128^2$; mean $\pm$ std over the
test set. Higher is more consistent. Best in \textbf{bold}, second best
\underline{underlined}.}
\label{tab:consistency}
\small
\begin{tabular}{lcc}
\toprule
Prior & Sparse-view CT & Accelerated MRI \\
\midrule
CNN  & $26.18 \pm 1.81$ & $33.34 \pm 2.43$ \\
FNO  & $\underline{31.39 \pm 1.53}$ & $38.33 \pm 1.68$ \\
AFNO & $30.92 \pm 1.58$ & $34.37 \pm 1.19$ \\
CNO  & $30.56 \pm 1.26$ & $\mathbf{42.62 \pm 2.94}$ \\
NNO  & $\mathbf{32.49 \pm 1.53}$ & $\underline{40.32 \pm 2.22}$ \\
\bottomrule
\end{tabular}
\end{table}

\begin{table}[t]
\centering
\caption{Grid-averaged PSNR (dB) / SSIM ($\times100$) under
single-resolution training on $128^2$ within two unrolled schemes,
averaged over $128^2$, $256^2$, and $512^2$. Best in \textbf{bold}, second
best \underline{underlined}.}
\label{tab:schemes}
\resizebox{\linewidth}{!}{%
\begin{tabular}{lcccc}
\toprule
& \multicolumn{2}{c}{Sparse-view CT} & \multicolumn{2}{c}{Accelerated MRI} \\
\cmidrule(lr){2-3}\cmidrule(lr){4-5}
Prior & ISTA-Net+ & MoDL & ISTA-Net+ & MoDL \\
\midrule
CNN
     & $28.12$ / $72.12$
     & $29.68$ / $70.61$
     & $32.35$ / $85.44$
     & $32.81$ / $86.23$ \\
FNO
     & $27.98$ / $76.89$
     & $27.29$ / $75.32$
     & $32.87$ / $87.87$
     & $32.96$ / $88.71$ \\
AFNO
     & $\underline{29.14}$ / $\underline{77.62}$
     & $\underline{30.00}$ / $\mathbf{82.94}$
     & $\underline{33.67}$ / $\mathbf{89.57}$
     & $33.78$ / $\underline{89.93}$ \\
CNO
     & $28.96$ / $69.25$
     & $29.54$ / $72.66$
     & $33.63$ / $88.47$
     & $\underline{33.84}$ / $86.49$ \\
NNO
     & $\mathbf{29.97}$ / $\mathbf{81.17}$
     & $\mathbf{30.15}$ / $\underline{82.65}$
     & $\mathbf{33.75}$ / $\underline{89.31}$
     & $\mathbf{34.47}$ / $\mathbf{90.36}$ \\
\bottomrule
\end{tabular}}
\end{table}

\subsection{Ablation}
\label{sec:exp_ablation}

Table~\ref{tab:ablation} changes one component of NNO at a time, using
ISTA-Net+ on sparse-view CT with single-resolution training on $128^2$.
Without the frequency filter ($M\equiv1$), the PSNR on $128^2$ decreases by
only 0.33~dB, but drops by 2.51 and 3.52~dB on $256^2$ and $512^2$. With a
linear spectral network $\widetilde\psi^{(\ell)}_{\theta}$, as in the Fourier
layers of FNO, the PSNR drops by 6.50~dB already on $128^2$; with a linear
spatial network $\phi^{(\ell)}_{\theta}$, it drops by 5.95~dB, and without the
residual connection in~\eqref{eq:nno_block_d}, by 9.22~dB. Sharing
$\theta_k$ across all stages reduces the number of parameters from 316.5K to
52.8K, with the same average PSNR and a higher average SSIM (84.09
vs.\ 81.17); this average PSNR is still higher than that of any other method
in Table~\ref{tab:ct_single}, including CTO-full with 172.5M parameters. All other experiments keep
separate $\theta_k$, as in ISTA-Net+.

\begin{table}[t]
\centering
\caption{Ablation of NNO with ISTA-Net+ on sparse-view CT under
single-resolution training on $128^2$. Entries are PSNR (dB) / SSIM
($\times100$), mean over the test set; the average is over the three grids.
All variants have 316.5K parameters, except the shared one (52.8K).
Best in \textbf{bold}.}
\label{tab:ablation}
\resizebox{\linewidth}{!}{%
\begin{tabular}{lcccc}
\toprule
Variant & $128^2$ & $256^2$ & $512^2$ & Average \\
\midrule
NNO
& $\mathbf{33.24}$ / $91.02$
& $28.92$ / $80.92$
& $27.76$ / $71.58$
& $\mathbf{29.97}$ / $81.17$ \\
w/o frequency filter
& $32.91$ / $91.27$
& $26.41$ / $63.03$
& $24.24$ / $53.22$
& $27.85$ / $69.17$ \\
Linear spectral network
& $26.74$ / $77.02$
& $26.88$ / $73.52$
& $26.32$ / $65.15$
& $26.65$ / $71.90$ \\
Linear spatial network
& $27.29$ / $77.59$
& $26.98$ / $66.85$
& $25.87$ / $52.55$
& $26.71$ / $65.66$ \\
w/o block residual
& $24.02$ / $51.41$
& $25.34$ / $48.32$
& $24.71$ / $42.15$
& $24.69$ / $47.29$ \\
\midrule
Shared across stages
& $32.96$ / $\mathbf{91.78}$
& $\mathbf{29.15}$ / $\mathbf{83.38}$
& $\mathbf{27.81}$ / $\mathbf{77.12}$
& $\mathbf{29.97}$ / $\mathbf{84.09}$ \\
\bottomrule
\end{tabular}}
\end{table}

\subsection{Mixed-Resolution Training}
\label{sec:exp_mixed}

We train each prior on its native reconstruction grids
$\mathcal T=\{128,256,512\}$ using fixed, nested supervision pools of
80, 40, and 20 images, respectively. For each grid, we average the
supervised loss over its available targets and combine the grid-wise
losses as in~\eqref{eq:training_loss_mix}. Theorem~\ref{thm:consistency}(a)
compares discretizations on common training pairs, while here the pools
also vary across grids (Appendix~\ref{app:exp_details}).
Table~\ref{tab:mixed} reports results on the same 40 held-out test images
at all three training resolutions and the unseen $384^2$ grid.

\paragraph{Sparse-view CT.}
NNO leads in PSNR and SSIM on all four grids. Compared with training on
$128^2$ alone, mixed-resolution supervision raises its PSNR by 2.78~dB at
both $256^2$ and $512^2$, with a decrease of 0.24~dB at $128^2$.
At $512^2$ and the unseen $384^2$, NNO leads the next-best PSNR by
0.69 and 0.72~dB. Averaged over the three training resolutions, NNO leads by
1.19~dB in PSNR and 4.26 points in SSIM ($\times100$).

\paragraph{Accelerated MRI.}
NNO has the highest PSNR and SSIM on every grid.
At $512^2$, its PSNR improves by 0.86~dB over single-resolution training
and exceeds that of AFNO by 0.10~dB; their SSIM scores there are 85.39
and 85.36 ($\times100$). At the unseen $384^2$, NNO's PSNR advantage is
0.14~dB.

\newcommand{\dmix}[2]{\,{\scriptsize\textcolor{gray}{($#1$\,/\,$#2$)}}}

\begin{table*}[t]
\centering
\caption{Mixed-resolution training with ISTA-Net+. Native-grid runs use
80, 40, and 20 training targets at $128^2$, $256^2$, and $512^2$,
respectively, in fixed nested pools. The \textit{interp.} rows are
fixed-grid controls trained on $512^2$, with lower-resolution targets
upsampled by Lanczos interpolation. Entries are mean PSNR (dB) / SSIM
($\times100$) over 40 test images, with changes from single-resolution
training in parentheses (Tables~\ref{tab:ct_single} and~\ref{tab:mri_single}).
The average is over $128^2$, $256^2$, and $512^2$. Best in \textbf{bold},
second best \underline{underlined}.}
\label{tab:mixed}
\small
\setlength{\tabcolsep}{5pt}
\resizebox{\textwidth}{!}{%
\begin{tabular}{lccccc}
\toprule
Prior & $128^2$ & $256^2$ & $512^2$ & $384^2$ (unseen) & Average \\
\midrule
\multicolumn{6}{l}{\emph{Sparse-view CT}} \\
CNN
& $30.61$ / $86.94$ \dmix{-1.57}{-3.35}
& $29.87$ / $79.00$ \dmix{+1.85}{+6.67}
& $28.13$ / $68.37$ \dmix{+3.98}{+14.63}
& $28.89$ / $73.10$
& $29.54$ / $78.10$ \dmix{+1.42}{+5.98} \\
CNN (interp.)
& $24.36$ / $66.12$ \dmix{-7.82}{-24.17}
& $26.66$ / $74.03$ \dmix{-1.36}{+1.70}
& $29.49$ / $\underline{77.86}$ \dmix{+5.34}{+24.12}
& --
& $26.84$ / $72.67$ \dmix{-1.28}{+0.55} \\
FNO
& $28.39$ / $82.02$ \dmix{-0.40}{+0.10}
& $28.24$ / $77.17$ \dmix{+0.21}{+0.63}
& $27.55$ / $73.28$ \dmix{+0.44}{+1.06}
& $27.85$ / $75.03$
& $28.06$ / $77.49$ \dmix{+0.08}{+0.60} \\
AFNO
& $31.35$ / $86.53$ \dmix{-0.11}{-2.06}
& $\underline{30.48}$ / $\underline{81.29}$ \dmix{+1.38}{+1.40}
& $\underline{29.85}$ / $76.85$ \dmix{+3.00}{+12.48}
& $\underline{30.28}$ / $\underline{79.19}$
& $\underline{30.56}$ / $\underline{81.56}$ \dmix{+1.42}{+3.94} \\
AFNO (interp.)
& $24.92$ / $57.29$ \dmix{-6.54}{-31.30}
& $26.22$ / $53.39$ \dmix{-2.88}{-26.50}
& $29.74$ / $73.69$ \dmix{+2.89}{+9.32}
& --
& $26.96$ / $61.46$ \dmix{-2.18}{-16.16} \\
CNO
& $\underline{31.36}$ / $\underline{90.62}$ \dmix{-2.06}{-1.81}
& $28.25$ / $63.75$ \dmix{+1.02}{+3.70}
& $27.80$ / $60.75$ \dmix{+1.58}{+5.47}
& $28.06$ / $62.37$
& $29.14$ / $71.71$ \dmix{+0.18}{+2.46} \\
NNO
& $\mathbf{33.00}$ / $\mathbf{91.22}$ \dmix{-0.24}{+0.20}
& $\mathbf{31.70}$ / $\mathbf{86.00}$ \dmix{+2.78}{+5.08}
& $\mathbf{30.54}$ / $\mathbf{80.24}$ \dmix{+2.78}{+8.66}
& $\mathbf{31.00}$ / $\mathbf{82.82}$
& $\mathbf{31.75}$ / $\mathbf{85.82}$ \dmix{+1.78}{+4.65} \\
\midrule
\multicolumn{6}{l}{\emph{Accelerated MRI}} \\
CNN
& $35.25$ / $94.71$ \dmix{-0.89}{-0.96}
& $33.27$ / $88.85$ \dmix{+1.77}{+3.54}
& $31.78$ / $83.43$ \dmix{+2.38}{+8.09}
& $32.30$ / $85.35$
& $33.43$ / $89.00$ \dmix{+1.08}{+3.56} \\
CNN (interp.)
& $28.01$ / $83.79$ \dmix{-8.13}{-11.88}
& $29.07$ / $81.31$ \dmix{-2.43}{-4.00}
& $29.91$ / $77.52$ \dmix{+0.51}{+2.18}
& --
& $29.00$ / $80.87$ \dmix{-3.35}{-4.57} \\
FNO
& $34.84$ / $94.35$ \dmix{+0.27}{+1.20}
& $33.20$ / $89.35$ \dmix{+0.38}{+1.36}
& $32.20$ / $85.08$ \dmix{+0.98}{+2.60}
& $32.48$ / $86.48$
& $33.41$ / $89.59$ \dmix{+0.54}{+1.72} \\
AFNO
& $34.99$ / $94.88$ \dmix{-1.37}{-1.08}
& $\underline{33.33}$ / $\underline{89.73}$ \dmix{-0.63}{-1.09}
& $\underline{32.38}$ / $\underline{85.36}$ \dmix{+1.69}{+3.43}
& $\underline{32.70}$ / $\underline{86.88}$
& $\underline{33.57}$ / $\underline{89.99}$ \dmix{-0.10}{+0.42} \\
AFNO (interp.)
& $30.05$ / $80.64$ \dmix{-6.31}{-15.32}
& $29.46$ / $73.95$ \dmix{-4.50}{-16.87}
& $31.16$ / $81.77$ \dmix{+0.47}{-0.16}
& --
& $30.22$ / $78.79$ \dmix{-3.45}{-10.78} \\
CNO
& $\underline{35.81}$ / $\underline{95.17}$ \dmix{-0.78}{-1.20}
& $32.67$ / $87.27$ \dmix{-0.11}{+0.21}
& $31.48$ / $81.80$ \dmix{-0.04}{-0.19}
& $31.81$ / $83.61$
& $33.32$ / $88.08$ \dmix{-0.31}{-0.39} \\
NNO
& $\mathbf{35.85}$ / $\mathbf{95.65}$ \dmix{-0.55}{+0.02}
& $\mathbf{33.66}$ / $\mathbf{90.49}$ \dmix{+0.42}{+1.42}
& $\mathbf{32.48}$ / $\mathbf{85.39}$ \dmix{+0.86}{+2.15}
& $\mathbf{32.84}$ / $\mathbf{87.24}$
& $\mathbf{34.00}$ / $\mathbf{90.51}$ \dmix{+0.25}{+1.20} \\
\bottomrule
\end{tabular}}
\end{table*}

\section{Conclusion}
\label{sec:conclusion}

In this paper, we formulate unrolled reconstruction in function space, with
a Nemytskii neural operator (NNO) serving as the prior. Since the parameters
are shared across grids and each data-consistency step uses the forward
operator discretized on the current grid, the network can be trained
directly on images at their native grids and can reconstruct the same
measurements on different grids. We prove that implementations of the
network on different grids approximate the same continuum network, and that
on common training pairs, the grid-wise losses approximate a shared
continuum loss up to constants independent of the network parameters.
On sparse-view CT and accelerated MRI, NNO trained on a single grid achieves
the highest average PSNR across grids among the priors. Mixed-resolution
training further improves its PSNR on the finest grid in both tasks, using
only 20 of the 80 training images as finest-grid targets. Our experiments
use simulated measurements, with images on coarser grids obtained by
downsampling; training on scans acquired at different resolutions is left
for future work.

\clearpage

\section*{LLM Usage Statement}
In this work, we used generative AI tools to assist with the proofs in Appendix~\ref{app:theory} (via Claude). AI was also used to write and review the code of our experiments (via ChatGPT and Grok) as follows: implementing model structures, assisting with hyperparameter tuning, writing scripts to launch training runs and to save and retrieve checkpoints and results, generating figures from experimental data. Additionally, we used generative AI tools to proofread and improve the writing of the manuscript (including correcting grammar mistakes, suggesting rephrasing to improve clarity and fluency, via Claude Code and ChatGPT). We have reviewed all AI-assisted work. All AI assistance was carried out under the direction of the authors, and the resulting material was subsequently checked and verified by the authors. We take responsibility for the final content of this work, including text, claims, or artifacts produced with the aid of generative AI.

 {\small
 \bibliographystyle{ieeenat_fullname}
 \bibliography{main}
 }


\clearpage
\appendix
\maketitlesupplementary

\section{Proofs for Section~\ref{sec:theory}}
\label{app:theory}

\subsection{Setting and Conventions}
\label{app:th_prelim}

We use the notation of Sections~\ref{sec:method} and~\ref{sec:theory},
with $\mathbb T^d=\mathbb R^d/\mathbb Z^d$, $h_N=1/N$ and
$\Gamma_N=\{x_j=h_Nj:\,j\in\{0,\dots,N-1\}^d\}$. The frequencies
representable on $\Gamma_N$ form
\[
    K_N=\big\{\kappa\in\mathbb Z^d:\,
    -\lfloor N/2\rfloor\le\kappa_i\le\lceil N/2\rceil-1\big\},
\]
and $K_N^\circ=\{\kappa\in\mathbb Z^d:\,|\kappa_i|<N/2\}$ is obtained by
removing the Nyquist frequencies $\kappa_i=-N/2$ of even $N$; for odd $N$,
$K_N^\circ=K_N$. The transform $a_N$ and the norm $\|\cdot\|_N$ are also
used for complex grid functions. Fields with several channels are treated
channel by channel, with $\ell^2$ norms over the channels. Since
$\mathcal B_{\mathrm{train}}$ is symmetric,
$\mathcal B_{\mathrm{train}}\subset K_N$ implies
$\mathcal B_{\mathrm{train}}\subset K_N^\circ$.

Because $\mathcal S_N$ is applied to the intermediate continuum fields, we
take these fields to be functions defined at every point of $\mathbb T^d$.
It suffices that $u^0$ and the updates $\mathcal G_k$ produce such
functions, since pointwise maps preserve this property and the spectral
steps add trigonometric polynomials (see~\eqref{eq:th_spectral} below).

\paragraph{Discrete Fourier facts.}
\begin{enumerate}
\item[(F1)] For every complex grid function $v$,
\begin{gather*}
    \|v\|_N^2=\sum_{\kappa\in K_N}|a_N[v](\kappa)|^2,\\
    v_j=\sum_{\kappa\in K_N}a_N[v](\kappa)e^{2\pi i\kappa\cdot x_j}.
\end{gather*}
If $v$ is real, then $a_N[v](\tilde\kappa)=\overline{a_N[v](\kappa)}$, where
$\tilde\kappa\in K_N$ is the representative of $-\kappa$ modulo $N$.
\item[(F2)] If $\sum_k|\widehat w(k)|<\infty$, then
$a_N[\mathcal S_Nw](\kappa)=\sum_{q\in\mathbb Z^d}\widehat w(\kappa+Nq)$
for $\kappa\in K_N$. In particular, a trigonometric polynomial $p$ with
modes in $K_N$ satisfies $a_N[\mathcal S_Np]=\widehat p$ on $K_N$ and
$\|\mathcal S_Np\|_N=\|p\|$.
\item[(F3)] If $w\in H^\beta(\mathbb T^d)$ with $\beta>d/2$ and
$F\subset K_N$, then
$a_{N,F}(w)\le C_{\beta,d}\,|F|^{1/2}N^{-\beta}\|w\|_{H^\beta}$.
\end{enumerate}
(F1) is Parseval's identity for the discrete Fourier transform. (F2)
follows by inserting the Fourier series of $w$ into $a_N$ and using
$h_N^d\sum_je^{2\pi i(k-\kappa)\cdot x_j}=1$ if $k\equiv\kappa$ (mod $N$)
and $0$ otherwise. For (F3), let $\kappa\in F$ and $q\ne0$. Then
$|\kappa+Nq|\ge|\kappa+Nq|_\infty\ge N(|q|_\infty-\tfrac12)
\ge\tfrac N2|q|_\infty$, so by (F2) and the Cauchy--Schwarz inequality
\[
\begin{aligned}
    &\big|a_N[\mathcal S_Nw](\kappa)-\widehat w(\kappa)\big|
    \le\sum_{q\ne0}|\widehat w(\kappa+Nq)|\\
    &\quad\le\Big(\sum_{q\ne0}\big(\tfrac N2|q|_\infty\big)^{-2\beta}
    \Big)^{1/2}\|w\|_{H^\beta},
\end{aligned}
\]
where the last sum is finite because $2\beta>d$. Summing the squares over
$\kappa\in F$ gives (F3).

\paragraph{Interpolation and band limitation.}
For $k\in\mathbb Z$ with $|k|<N/2$ let $b_k(t)=e^{2\pi ikt}$, and for even
$N$ let $b_{-N/2}(t)=\cos(\pi Nt)+\sin(\pi Nt)$. For $\kappa\in K_N$ put
$b_\kappa(x)=\prod_{i=1}^db_{\kappa_i}(x_i)$, and for grid functions $v$
define
\[
    I_Nv=\sum_{\kappa\in K_N}a_N[v](\kappa)\,b_\kappa .
\]
The functions $b_k$ are orthonormal in $L^2(\mathbb T)$, since
$b_{-N/2}$ has the Fourier coefficients $\frac12\mp\frac i2$ at $\pm N/2$.
Hence the $b_\kappa$ are orthonormal in $L^2(\mathbb T^d)$, and
$b_\kappa(x_j)=e^{2\pi i\kappa\cdot x_j}$. With (F1) this gives:
\begin{enumerate}
\item[(F4)] $I_N$ is an isometry from $(\mathbb R^{N^d},\|\cdot\|_N)$ into
$L^2(\mathbb T^d)$, and $\mathcal S_NI_Nv=v$. Since
$b_{\tilde\kappa}=\overline{b_\kappa}$, $I_Nv$ is real for real $v$.
\end{enumerate}
Let $V_N=I_N(\mathbb R^{N^d})$ and let $T_N$ be the orthogonal projection
of $\mathcal X$ onto $V_N$. Then $I_N\mathcal S_Np=p$ for $p\in V_N$, and by
(F2) $V_N$ contains every real trigonometric polynomial with modes in
$K_N^\circ$. For odd $N$, $I_N$ is the usual trigonometric interpolation and
$T_N$ removes the modes outside $K_N$. For even $N$, $\cos(\pi Nt)$ and
$\sin(\pi Nt)$ cannot both be represented, since the samples of the latter
vanish; in each Nyquist direction $T_N$ keeps the component along
$b_{-N/2}$, which has the same samples as $\cos(\pi Nt)$. This choice makes
$I_N$ an isometry for every $N$.

\subsection{Two Lemmas}
\label{app:th_lemmas}

\begin{lemma}[Rectangle rule for fields with interfaces]
\label{lem:th_quadrature}
Let $\Sigma\subset\mathbb T^d$ be a finite union of compact $C^1$
hypersurfaces, and let $w$ be bounded with
$|w(x)-w(x')|\le L_w|x-x'|$ whenever the segment from $x$ to $x'$ does not
meet $\Sigma$. There is $C_\Sigma$ such that, for every $N$, every
$\kappa\in\mathbb Z^d$, and whatever the values of $w$ on $\Sigma$,
\[
\begin{aligned}
    &\big|a_N[\mathcal S_Nw](\kappa)-\widehat w(\kappa)\big|\\
    &\quad\le\Big(\sqrt d\,\big(L_w+2\pi|\kappa|\,\|w\|_\infty\big)
    +2C_\Sigma\|w\|_\infty\Big)N^{-1}.
\end{aligned}
\]
Hence $a_{N,F}(w)\le C_{F,\Sigma}(L_w+\|w\|_\infty)N^{-1}$ for every
finite $F\subset K_N$.
\end{lemma}

\begin{proof}
Let $Q_j=x_j+[0,h_N)^d$ and $g(x)=w(x)e^{-2\pi i\kappa\cdot x}$. Then
$a_N[\mathcal S_Nw](\kappa)-\widehat w(\kappa)
=\sum_j\big(h_N^dg(x_j)-\int_{Q_j}g\big)$. If $\overline{Q_j}$ does not
meet $\Sigma$, then $|g(x)-g(x_j)|\le(L_w+2\pi|\kappa|\|w\|_\infty)
\sqrt d\,h_N$ for $x\in Q_j$, so the term is at most
$h_N^d\sqrt d\,(L_w+2\pi|\kappa|\|w\|_\infty)h_N$; there are at most $N^d$
such cells. Every other term is at most $2\|w\|_\infty h_N^d$. A cell
whose closure meets $\Sigma$ lies in the $\sqrt d\,h_N$-neighborhood of
$\Sigma$. Since $\Sigma$ is a finite union of compact $C^1$ hypersurfaces,
this neighborhood has volume at most $C_\Sigma h_N$ for every $N$, so there
are at most $C_\Sigma N^{d-1}$ such cells. Adding the two contributions
gives the first bound; the second follows by summing the squares over
$\kappa\in F$.
\end{proof}

\begin{lemma}[Error propagation]
\label{lem:th_propagation}
Let a computation produce continuum fields $z_1,\dots,z_J$ and grid fields
$z_1^N,\dots,z_J^N$ in this order, and let
$e_t=\|z_t^N-\mathcal S_Nz_t\|_N$. Suppose that for each $t$ there are a
set $\mathrm{pa}(t)\subset\{1,\dots,t-1\}$, constants $c_{t,s}\ge0$ that do
not depend on $N$, and $\sigma_t\ge0$ such that
$e_t\le\sum_{s\in\mathrm{pa}(t)}c_{t,s}e_s+\sigma_t$. Then
\[
    e_t\le\sum_{r\le t}\omega_{t,r}\,\sigma_r,
\]
where $\omega_{t,t}=1$,
$\omega_{t,r}=\sum_{s\in\mathrm{pa}(t)}c_{t,s}\,\omega_{s,r}$ for $r<t$,
and $\omega_{s,r}=0$ for $r>s$. The coefficients $\omega_{t,r}\ge0$ do not
depend on $N$.
\end{lemma}

\begin{proof}
Induction on $t$: insert the bounds for $e_s$, $s\in\mathrm{pa}(t)$, into the
inequality for $e_t$.
\end{proof}

\subsection{Proof of Theorem~\ref{thm:local}}
\label{app:th_proof_local}

\emph{(a)} The $K$ stages consist of data-consistency updates, the
pointwise maps $\gamma_{\theta_k}$, $\phi^{(\ell)}_{\theta_k}$,
$\xi_{\theta_k}$, the spectral steps
\eqref{eq:nno_block_a}--\eqref{eq:nno_block_c}, and the residual additions
in~\eqref{eq:nno_block_d} and~\eqref{eq:nno_prior}. We list the fields in
the order of computation and bound the node-wise error of the output of each
operation by the errors of its inputs.

\emph{Initialization.} The grid uses $\mathcal S_Nu^0$, so the error is
zero.

\emph{Pointwise maps.} Let $\Phi$ be one of $\gamma_{\theta_k}$,
$\phi^{(\ell)}_{\theta_k}$, $\xi_{\theta_k}$, and let $L_\Phi$ bound its
derivative in the field variable. For a grid field $g$ and a continuum
field $z$, $\Phi(x_j,(\mathcal S_Nz)_j)=(\mathcal S_N\Phi(\cdot,z))_j$ at
every node, so
$\|\Phi(\cdot,g)-\mathcal S_N\Phi(\cdot,z)\|_N\le L_\Phi\|g-\mathcal S_Nz\|_N$.

\emph{Residual additions.} The error of a sum is at most the sum of the
errors of its terms.

\emph{Spectral steps.} Let $w$ be the continuum field entering
\eqref{eq:nno_block_a}, $g$ its grid counterpart,
$e_w=\|g-\mathcal S_Nw\|_N$, $\psi=\psi^{(\ell)}_{\theta_k}$, and $L_\psi$
the largest Lipschitz constant of $\psi(\kappa,\cdot)$ over the finite set
$\mathcal B_{\mathrm{train}}$. Since $\psi(\kappa,\cdot)$ is the identity
for $\kappa\notin\mathcal B_{\mathrm{train}}$ and $w$ is real, the
continuum step gives
\begin{equation}
    \operatorname{Re}\mathcal F^{-1}\psi(\cdot,\widehat w)
    =w+\operatorname{Re}R,
    \label{eq:th_spectral}
\end{equation}
where $R=\sum_{\kappa\in\mathcal B_{\mathrm{train}}}\widehat R(\kappa)\,
e^{2\pi i\kappa\cdot x}$ and
$\widehat R(\kappa)=\psi(\kappa,\widehat w(\kappa))-\widehat w(\kappa)$.
By (F1), the grid step gives $g+\operatorname{Re}\mathcal S_NR_N$, where
$R_N$ is defined as $R$ with $\widehat w$ replaced by $a_N[g]$. Since
$R_N-R$ has modes in $\mathcal B_{\mathrm{train}}\subset K_N$, (F2) and
$|\operatorname{Re}z|\le|z|$ give
\[
\begin{aligned}
    &\big\|g+\operatorname{Re}\mathcal S_NR_N
      -\mathcal S_N(w+\operatorname{Re}R)\big\|_N\\
    &\quad\le e_w+\|\widehat R_N-\widehat R\|_{\ell^2(\mathcal B_{\mathrm{train}})}.
\end{aligned}
\]
Moreover, $|\widehat R_N(\kappa)-\widehat R(\kappa)|\le(1+L_\psi)
|a_N[g](\kappa)-\widehat w(\kappa)|$, and by (F1)
\[
\begin{aligned}
    &\|a_N[g]-\widehat w\|_{\ell^2(\mathcal B_{\mathrm{train}})}\\
    &\quad\le\|a_N[g-\mathcal S_Nw]\|_{\ell^2(K_N)}
      +a_{N,\mathcal B_{\mathrm{train}}}(w)\\
    &\quad=e_w+a_{N,\mathcal B_{\mathrm{train}}}(w).
\end{aligned}
\]
The error of the output of the spectral step is therefore at most
$(2+L_\psi)\,e_w+(1+L_\psi)\,a_{N,\mathcal B_{\mathrm{train}}}(w)$.

\emph{Data-consistency updates.} Let $L_{\mathcal G}$ bound the Lipschitz
constants of the $\mathcal G_{k,N}$ uniformly in $N$, let $g$ be the grid
iterate and $u^k$ the continuum iterate. Then
\[
\begin{aligned}
    &\|\mathcal G_{k,N}(g)-\mathcal S_N\mathcal G_k(u^k)\|_N\\
    &\quad\le\|\mathcal G_{k,N}(g)-\mathcal G_{k,N}(\mathcal S_Nu^k)\|_N
      +\varepsilon^{\mathrm{data}}_k(N)\\
    &\quad\le L_{\mathcal G}\,\|g-\mathcal S_Nu^k\|_N
      +\varepsilon^{\mathrm{data}}_k(N).
\end{aligned}
\]

Every operation thus satisfies the hypothesis of
Lemma~\ref{lem:th_propagation}, with constants
$c_{t,s}\in\{1,L_\Phi,2+L_\psi,L_{\mathcal G}\}$ that do not depend on $N$
and source terms $\sigma_t\in\{0,\varepsilon^{\mathrm{data}}_k(N),
(1+L_\psi)\,a_{N,\mathcal B_{\mathrm{train}}}(w)\}$. Applied to the output
$u^K$, the lemma gives~\eqref{eq:th_localization} with
$\Lambda'_k=\omega_{J,t_k}$ and $\Lambda_w=(1+L_\psi)\,\omega_{J,t_w}$, where
$t_k$ indexes the data-consistency update of stage $k$ and $t_w$ the
spectral step whose input is $w$.

\emph{(b)} The spectral terms are
$\Lambda_w\,a_{N,\mathcal B_{\mathrm{train}}}(w)$, where neither the
continuum fields $w$ nor the constants $\Lambda_w$ depend on $N$, and
$\mathcal B_{\mathrm{train}}$ is a fixed finite set. If
$w\in C^\infty(\mathbb T^d)$, then $w\in H^\beta$ for every $\beta$, and
(F3) gives $O(N^{-\beta})$ for every $\beta>0$. If $w$ is piecewise
Lipschitz with finitely many $C^1$ interfaces,
Lemma~\ref{lem:th_quadrature} gives $O(N^{-1})$. \qed

\subsection{Proof of Corollary~\ref{cor:exact}}
\label{app:th_proof_cor}

We work under the assumptions of Theorem~\ref{thm:local}. The measurements
are $\mathcal Au=(\widehat u(\kappa))_{\kappa\in\mathcal P}$ with the inner
product $\langle r,s\rangle=\operatorname{Re}\sum_{\kappa\in\mathcal P}
r_\kappa\overline{s_\kappa}$. Hence, for real $u$,
\[
    \mathcal A^*r=\operatorname{Re}\sum_{\kappa\in\mathcal P}
    r_\kappa e^{2\pi i\kappa\cdot x},
    \qquad
    \widehat{\mathcal A^*\mathcal Au}=\mu\,\widehat u,
\]
with $\mu=\frac12(\mathbf 1_{\mathcal P}+\mathbf 1_{-\mathcal P})$. The
same computation with $\langle\cdot,\cdot\rangle_N$ gives
$\mathcal A_N^*r=\mathcal S_N\mathcal A^*r$, and by (F1)
$\mathcal A_N^*\mathcal A_N$ multiplies $a_N[v]$ by the same $\mu$.

Let $S=\{0\}\cup\mathcal P\cup(-\mathcal P)\cup\mathcal B_{\mathrm{train}}$
and let $V_S$ be the real trigonometric polynomials with modes in $S$. Since
$\pm\mathcal P$ and $\mathcal B_{\mathrm{train}}$ are symmetric subsets of
$K_N$, $S\subset K_N^\circ$. By (F2), for $p,p'\in V_S$,
$a_N[\mathcal S_Np]=\widehat p$ on $K_N$ and
$\langle\mathcal S_Np,\mathcal S_Np'\rangle_N=\langle p,p'\rangle$. We show
by induction over the operations that every continuum field lies in $V_S$
and that every grid field equals the samples of its continuum counterpart.
\begin{itemize}
\item The initialization is $u^0=\mathcal A^*y\in V_S$, and the grid uses
$\mathcal A_N^*y=\mathcal S_Nu^0$.
\item Affine, position-independent maps $s\mapsto Ws+b$ act on Fourier
coefficients as $\widehat z\mapsto W\widehat z+b\,\mathbf 1_{\{0\}}$, so
they preserve $V_S$; being pointwise, they commute with $\mathcal S_N$.
\item For $w\in V_S$, the grid spectral step receives
$a_N[\mathcal S_Nw]=\widehat w$, so $R_N=R$ and it returns the samples
of~\eqref{eq:th_spectral}, which lies in $V_S$ because $R$ has modes in
$\mathcal B_{\mathrm{train}}$.
\item For $u\in V_S$, $\mathcal A_N\mathcal S_Nu=\mathcal Au$ and
$\mathcal A_N^*r=\mathcal S_N\mathcal A^*r$. Hence the gradient update
$u-\tau\,(\mathcal A^*\mathcal Au-\mathcal A^*y)$ and the proximal update
$(I+\lambda\mathcal A^*\mathcal A)^{-1}(u+\lambda\mathcal A^*y)$, which
applies the multiplier $(1+\lambda\mu)^{-1}$ to $u+\lambda\mathcal A^*y$,
map $V_S$ into $V_S$ and commute with $\mathcal S_N$.
\item Residual additions are linear.
\end{itemize}
Hence $u^K_N(y;\Theta)=\mathcal S_Nu^K(y;\Theta)$.

For the second claim, fix $\kappa\notin\pm\mathcal P$ and follow the
Fourier coefficients of every field at $\kappa$ and $-\kappa$. Each
operation computes them from the coefficients of its inputs at $\kappa$ and
$-\kappa$ alone: the affine maps act frequency by frequency, the spectral
step applies $\psi(\pm\kappa,\cdot)$ and its real part couples only
$\kappa$ and $-\kappa$, and the data-consistency updates leave these
coefficients unchanged, because $\mu(\pm\kappa)=0$ and $\mathcal A^*y$ has
no modes outside $\pm\mathcal P$. The coefficients of $u^0$ at $\pm\kappa$
vanish. The coefficients of $u^K$ at $\pm\kappa$ are therefore obtained
from zero by maps that do not involve $y$. By the first claim, the same
holds for $a_N[u^K_N]$ on $K_N\setminus(\pm\mathcal P)$. \qed

If the proximal map is computed by a fixed number of conjugate-gradient
iterations, the first claim still holds, because these iterations use only
$\mathcal A^*\mathcal A$, linear combinations, and inner products, all of
which commute with $\mathcal S_N$ on $V_S$. The second claim may then fail,
since the step sizes are inner products that depend on $y$.

\subsection{Proof of Theorem~\ref{thm:consistency}}
\label{app:th_proof_loss}

\begin{lemma}
\label{lem:grid_loss}
For every $\Theta$ and $N$,
\[
    \mathcal L_N(\Theta)+\widehat\varepsilon_N^2
    =\frac1n\sum_{i=1}^n
    \big\|I_Nu^K_N(y_i;\Theta)-u_i^\dagger\big\|^2 .
\]
\end{lemma}

\begin{proof}
Since $T_Nu_i^\dagger\in V_N$, (F4) gives
$I_Nu^\dagger_{i,N}=I_N\mathcal S_NT_Nu_i^\dagger=T_Nu_i^\dagger$ and
$\|u^K_N(y_i;\Theta)-u^\dagger_{i,N}\|_N
=\|I_Nu^K_N(y_i;\Theta)-T_Nu_i^\dagger\|$. The function
$I_Nu^K_N(y_i;\Theta)-T_Nu_i^\dagger$ lies in $V_N$ and is therefore
orthogonal to $(I-T_N)u_i^\dagger$. Pythagoras' theorem, averaged over $i$,
gives the identity.
\end{proof}

\emph{Bound on $\delta_N$.} Write $u_i=u^K(y_i;\Theta)$ and
$u_{i,N}=u^K_N(y_i;\Theta)$. By (F4),
\[
\begin{aligned}
    &\|I_Nu_{i,N}-u_i\|\\
    &\quad\le\|I_N(u_{i,N}-\mathcal S_Nu_i)\|+\|(I-I_N\mathcal S_N)u_i\|\\
    &\quad=\|u_{i,N}-\mathcal S_Nu_i\|_N+\|(I-I_N\mathcal S_N)u_i\|,
\end{aligned}
\]
and Minkowski's inequality over $i$ gives the bound on $\delta_N$ stated in
Section~\ref{sec:mixed_training}. Its first term is the root-mean-square of
the node-wise errors bounded in Theorem~\ref{thm:local}. This bound is the
only place where the assumptions of Theorem~\ref{thm:local} enter; parts
(a) and (b) below hold for every $\Theta$ and $N$.

\emph{(a)} Let $a=(\mathcal L_N+\widehat\varepsilon_N^2)^{1/2}$ and
$b=\mathcal L^{1/2}$. By Lemma~\ref{lem:grid_loss}, $a$ and $b$ are the
norms of $(I_Nu_{i,N}-u_i^\dagger)_i$ and $(u_i-u_i^\dagger)_i$ in
$\mathcal X^n$ with the norm $\big(\frac1n\sum_i\|\cdot\|^2\big)^{1/2}$, so
the triangle inequality gives $|a-b|\le\delta_N$. Hence
$|a^2-b^2|=|a-b|\,(a+b)\le\delta_N\,(2b+\delta_N)$, which is the first
bound. Since $\sum_{N\in\mathcal T}\pi_N=1$,
\[
\begin{aligned}
    &\mathcal L_{\mathrm{mix}}
    +\sum_{N\in\mathcal T}\pi_N\widehat\varepsilon_N^2-\mathcal L\\
    &\quad=\sum_{N\in\mathcal T}\pi_N\big(\mathcal L_N
    +\widehat\varepsilon_N^2-\mathcal L\big),
\end{aligned}
\]
and the first bound, applied to each term, gives the second.

\emph{(b)} By the proof of (a), $(\mathcal L_N+\widehat\varepsilon_N^2)^{1/2}$
and $(\mathcal L_{N'}+\widehat\varepsilon_{N'}^2)^{1/2}$ lie within
$\delta_N$ and $\delta_{N'}$ of $\mathcal L^{1/2}$, respectively. The
argument does not require $N$ or $N'$ to be a training resolution. \qed

\emph{Minimizers.} Suppose that
$|\mathcal L_N+\widehat\varepsilon_N^2-\mathcal L|\le\eta$ for all $\Theta$
in a parameter set and that $\Theta_N$ minimizes $\mathcal L_N$ on this set.
Since $\widehat\varepsilon_N$ does not depend on $\Theta$, for every
$\Theta$ in the set
\[
\begin{aligned}
    \mathcal L(\Theta_N)
    &\le\mathcal L_N(\Theta_N)+\widehat\varepsilon_N^2+\eta\\
    &\le\mathcal L_N(\Theta)+\widehat\varepsilon_N^2+\eta
    \le\mathcal L(\Theta)+2\eta,
\end{aligned}
\]
so $\Theta_N$ minimizes $\mathcal L$ up to $2\eta$. The same argument
applies to $\mathcal L_{\mathrm{mix}}$. This makes precise the statement
after Theorem~\ref{thm:consistency}.


\section{Experimental Details}
\label{app:exp_details}

\subsection{Data}
\label{app:data}

\paragraph{Sparse-view CT.}
We use 120 slices of the SIIM CT dataset~\cite{mader2016ctmedicalimages}
(from TCGA-LUAD~\cite{albertina2016tcgaluad,clark2013tcia}). The DICOM
values are converted to Hounsfield units with the rescale slope and
intercept, clipped to $[-1000,1000]$~HU, and normalized to $[0,1]$. The 80
training slices and the 40 test slices have no patient in common, and the
40 test slices come from 40 different patients.

\paragraph{Accelerated MRI.}
We use the qDESS scans of SKM-TEA~\cite{desai2021skm}, combine the two
echoes by root-sum-of-squares, and take one slice from each of 120 scans.
The 80 training scans and the 40 test scans have no patient in common, and
the 40 test slices come from 40 different patients. Images are clipped to
the 0.5th--99.5th percentile window estimated on the training set and
normalized to $[0,1]$.

In both tasks, the images have $512^2$ pixels and are resized to $128^2$,
$256^2$ and $384^2$ by Lanczos interpolation. The $384^2$ grid is used only
for evaluation.

\subsection{Measurements and Forward Operators}
\label{app:forward}

All measurements are simulated from the $512^2$ images without noise, and
every grid uses the same $y$. The initialization $u^0$ is FBP for CT and
zero filling for MRI.

\paragraph{Sparse-view CT.}
Sinograms are simulated in a parallel-beam geometry with 720 views and 300
detector bins, of which 36 uniformly spaced views are kept. On a grid of
size $N$, $\mathcal A_N$ computes these projections from the $N\times N$
image: the field of view, the projection angles, the detector span and the
detector bins do not depend on $N$, and the image is sampled along each ray
and summed with the physical step size.

\paragraph{Accelerated MRI.}
The forward operator is $\mathcal A_N=P_\Omega F_{512}R_{512\leftarrow N}$,
where $R_{512\leftarrow N}$ resamples the image bilinearly to the $512^2$
measurement grid ($R_{512\leftarrow512}=I$), $F_{512}$ is the
two-dimensional DFT, and $P_\Omega$ applies a fixed one-dimensional
Cartesian mask that keeps 128 of the 512 phase-encoding lines, including
the 41 central lines.

In both tasks, the adjoint $\mathcal A_N^*$ is computed exactly as a
vector--Jacobian product, and the data-consistency gradient is divided by
an estimate of the squared operator norm computed with 8 power iterations;
the normalized gradient is therefore the same for the inner product
$\langle\cdot,\cdot\rangle_N$ of Section~\ref{sec:disc_train}.

\subsection{Unrolled Schemes and Priors}
\label{app:priors}

ISTA-Net+ uses $K=6$ stages for CT and $K=8$ for MRI, with separate prior
parameters in each stage; each stage takes a gradient step with a learnable
step size before the prior. MoDL uses one prior shared by 4 stages, and its
proximal data-consistency step is computed by 4 conjugate-gradient
iterations. Table~\ref{tab:param_counts} lists the parameter counts.

\begin{table}[t]
\centering
\caption{Number of trainable parameters of each prior and of the
task-specific methods.}
\label{tab:param_counts}
\small
\setlength{\tabcolsep}{5pt}
\begin{tabular}{lcccc}
\toprule
& \multicolumn{2}{c}{ISTA-Net+} & \multicolumn{2}{c}{MoDL} \\
\cmidrule(lr){2-3}\cmidrule(lr){4-5}
Prior & CT & MRI & CT & MRI \\
\midrule
CNN  & 317.2K & 199.5K & 50.1K & 85.2K \\
FNO  & 393.8K & 321.8K & 49.2K & 79.9K \\
AFNO & 407.8K & 422.4K & 51.2K & 78.5K \\
CNO  & 316.7K & 422.3K & 52.8K & 79.5K \\
NNO  & 316.5K & 221.3K & 52.8K & 73.8K \\
\midrule
\multicolumn{5}{l}{CT: CTO 313.7K, CTO-full 172.5M} \\
\multicolumn{5}{l}{MRI: NO-VarNet 221.6K, NO-VarNet-full 139.7M} \\
\bottomrule
\end{tabular}
\end{table}

\paragraph{NNO.}
Within ISTA-Net+, NNO uses $c=32$ channels and $L=2$ blocks per stage for
CT, and $c=24$ and $L=1$ for MRI; within MoDL, it uses $c=32$ with $L=2$
(CT) and $L=3$ (MRI). The lifting and aggregation networks are pointwise
MLPs with hidden widths 64 and 48; they and the spatial networks take the
coordinates as additional inputs. The spectral and spatial networks are pointwise MLPs with three
linear layers and hidden width 64, with GELU after the first two layers and
no activation after the last. The frequency filter uses
$\kappa_{\mathrm{train}}^{\max}=64$, the Nyquist frequency of the $128^2$
grid, for both single- and mixed-resolution training; hence
$\mathcal B_{\mathrm{train}}\subset K_N$ for all $N\ge128$.

\paragraph{FNO.}
FNO replaces the spatial--spectral blocks of NNO by residual Fourier layers
and keeps its lifting and aggregation networks. Within ISTA-Net+, it has
width 4, depth 4 and $16\times16$ retained modes for CT, and width 4, depth
2 and $16\times16$ modes for MRI; within MoDL, width 4, depth 3 and
$16\times16$ modes for CT, and width 6, depth 5 and $10\times10$ modes for
MRI.

\paragraph{AFNO.}
AFNO blocks with eight block-diagonal groups act on the pixel tokens of the
native grid, without patch embedding or positional encoding. Lifting and
projection follow NNO, with normalized coordinates in $[-1,1]^2$ as
additional inputs, and a learnable residual scale is used. The same
block-diagonal complex MLP is applied to every retained Fourier mode. As in
the original AFNO~\cite{guibas2022afno}, the Fourier transform is
orthonormal and the number of retained modes is a fixed fraction of the
grid size, so the prior changes with the grid. Within ISTA-Net+, AFNO uses
4 blocks with embedding dimension 40 and MLP ratio 4 for CT, and 2 blocks
with dimension 48 and ratio 4 for MRI; within MoDL, 4 blocks with dimension
32 and ratio 4.5 for CT, and 8 blocks with dimension 16 and ratio 16 for
MRI.

\paragraph{CNO.}
CNO~\cite{raonic2023cno} is a residual U-shaped operator with two
resolution levels, four residual blocks per level, seven bottleneck residual
blocks and a channel multiplier of 8, defined on the training grid $128^2$.
On a grid of size $N$, the prior update is
$z+\beta\,\mathcal I_{N\leftarrow128}[G(\mathcal P_{128\leftarrow N}z)]$,
where $\mathcal P_{128\leftarrow N}$ truncates the spectrum to $128^2$,
$G$ is the CNO, $\mathcal I_{N\leftarrow128}$ zero-pads the spectrum back
to $N^2$, and $\beta$ is learnable. Data consistency and the residual
connection act on the native grid.

\paragraph{CNN.}
Within ISTA-Net+, the CNN prior follows the original
design~\cite{zhang2018ista}: a convolutional forward transform,
soft-thresholding with a learnable threshold, and a symmetric backward
transform, with 38 (CT) and 26 (MRI) feature channels and the symmetry loss
with weight 0.01. Within MoDL~\cite{aggarwal2019modl}, the prior is a
residual CNN with batch normalization and four layers of 52 channels (CT)
or three layers of 96 channels (MRI), and the regularization weight is
learnable and initialized to 0.2. All kernels are $3\times3$.

\paragraph{CTO.}
We adapt the official CTO~\cite{datta2025cto} to our fixed-observation
setting. Two averaged residual branches in the sinogram and
detector-frequency domains preprocess the sinogram; three cascades then
alternate FBP, an image-domain UDNO and forward projection, and the final
image is obtained by FBP. CTO uses width 12 and one pooling level per
domain, and CTO-full uses width 32 and four pooling levels. Both use the
DISCO basis $(6,7)$ and base radius cutoffs of 0.02 (sinogram) and 0.0625
(image).

\paragraph{NO-VarNet.}
We adapt the official NO-VarNet~\cite{jatyani2025unified} to single-coil
MRI with fixed $512^2$ measurements. A k-space neural operator first
inpaints the undersampled k-space with an additive skip connection; the
result is transformed to the image domain, bilinearly resized to the target
grid, and converted to a magnitude image. Image-domain cascades then
alternate data-consistency updates with a UDNO prior. NO-VarNet uses 5
cascades of width 10 with one pooling level and a k-space operator of width
2, and NO-VarNet-full uses 12 cascades of width 18 with four pooling levels
and a k-space operator of width 16. Both use the DISCO basis $(6,7)$ with
radius cutoff 0.02; the cascade kernels are built on the training grid
$128^2$ and reused on the other grids, as in the official implementation.

\subsection{Training and Evaluation}
\label{app:training}

All models are trained for 100 epochs with the mean-squared error, Adam
with weight decay $10^{-7}$, and gradient clipping at 1.0.
Single-resolution training uses $128^2$ and batch size 2. NNO and FNO
within ISTA-Net+ start with learning rate $5\times10^{-3}$, which is
multiplied by 0.7 after 40\% of training; all other models use a fixed
learning rate of $5\times10^{-3}$. The schedules were chosen per model.
Each task runs on one NVIDIA A100 GPU
with eight CPU cores and 80~GB of host memory. All results are computed on
the 40 test slices of each task. PSNR and SSIM are computed after clipping
the reconstruction to $[0,1]$, the range of the targets, with data range 1;
SSIM uses a $7\times7$ uniform window with reflect padding on every grid.
Clipping is applied only for evaluation.

\subsection{Mixed-Resolution Training}
\label{app:mixed}

\paragraph{Availability of training targets.}
For each task, let $S_{128}$ be the 80 training slices. We fix a subset
$S_{256}\subset S_{128}$ of size 40 and a subset $S_{512}\subset S_{256}$
of size 20 before training. Thus 40 slices supply targets on the coarsest
grid only, 20 on the two coarser grids, and 20 on all three grids.
At resolution $N$, the empirical mean in~\eqref{eq:training_loss_N} is
taken over $S_N$ for this experiment. The measurements remain the ones
simulated on the common $512^2$ measurement grid as described in
Section~\ref{sec:exp_setup}. All four reconstruction resolutions are
evaluated on the same 40 held-out slices per task.

\paragraph{Grids in each mini-batch.}
The parameters are shared across grids, training runs for 100 epochs, and
the $384^2$ grid is never used for training. Each optimizer step uses an
effective batch of 4 image--grid pairs, with at most 4, 2 and 1 samples from
$128^2$, $256^2$ and $512^2$, respectively. The samples of each grid are
processed together, and the losses of the grids are weighted by their
numbers of samples, so $\pi_N$ in~\eqref{eq:training_loss_mix} is the
expected fraction of samples from grid $N$. Each epoch draws 80
image--grid pairs; for each pair, the grid $N$ is drawn with probability
$\pi_N$ and the image from $S_N$. The probabilities
$(\pi_{128},\pi_{256},\pi_{512})$ are $(0.9,0.1,0)$ for the first 5 epochs,
change linearly to $(0.6,0.3,0.1)$ over the next 20 epochs, and remain at
$(0.6,0.3,0.1)$ for the last 75 epochs.

\paragraph{Fixed-grid controls.}
CNN (interp.) and AFNO (interp.) use the same image--grid pairs as the
native-grid runs, but every target is upsampled to $512^2$ by Lanczos
interpolation, and reconstruction and data consistency take place on
$512^2$.

\section{Additional Results}
\label{app:exp_results}

\subsection{Cross-Grid Consistency}
\label{app:consistency}

Table~\ref{tab:consistency_all} gives the cross-grid PSNR for every pair of
the three grids; Table~\ref{tab:consistency} shows the pair $128^2$ and
$512^2$. The finer reconstruction is downsampled to the coarser grid with
the floating-point Lanczos filter of PIL and clipped to $[0,1]$, and the
PSNR uses the data range 1. No ground truth is involved.

\begin{table}[t]
\centering
\caption{Cross-grid PSNR (dB) between the reconstructions on two grids
under single-resolution training on $128^2$ with ISTA-Net+; mean $\pm$ std
over the test set. Best in \textbf{bold}, second best
\underline{underlined}.}
\label{tab:consistency_all}
\footnotesize
\setlength{\tabcolsep}{3pt}
\begin{tabular}{lccc}
\toprule
Prior & $128^2$--$256^2$ & $256^2$--$512^2$ & $128^2$--$512^2$ \\
\midrule
\multicolumn{4}{l}{\emph{Sparse-view CT}} \\
CNN & $30.60 \pm 2.20$ & $27.53 \pm 1.83$ & $26.18 \pm 1.81$ \\
FNO & $33.57 \pm 1.68$ & $\mathbf{39.78 \pm 1.54}$ & $\underline{31.39 \pm 1.53}$ \\
AFNO & $\underline{33.97 \pm 1.61}$ & $35.90 \pm 1.76$ & $30.92 \pm 1.58$ \\
CNO & $32.76 \pm 1.40$ & $35.37 \pm 1.66$ & $30.56 \pm 1.26$ \\
NNO & $\mathbf{34.22 \pm 1.34}$ & $\underline{38.10 \pm 1.98}$ & $\mathbf{32.49 \pm 1.53}$ \\
\midrule
\multicolumn{4}{l}{\emph{Accelerated MRI}} \\
CNN & $35.53 \pm 2.76$ & $38.04 \pm 2.44$ & $33.34 \pm 2.43$ \\
FNO & $\underline{44.62 \pm 3.20}$ & $41.19 \pm 1.34$ & $38.33 \pm 1.68$ \\
AFNO & $42.34 \pm 2.69$ & $36.63 \pm 1.10$ & $34.37 \pm 1.19$ \\
CNO & $\mathbf{45.57 \pm 3.44}$ & $\underline{44.57 \pm 2.61}$ & $\mathbf{42.62 \pm 2.94}$ \\
NNO & $43.99 \pm 2.62$ & $\mathbf{44.61 \pm 1.85}$ & $\underline{40.32 \pm 2.22}$ \\
\bottomrule
\end{tabular}
\end{table}

\subsection{MoDL Results on Each Grid}
\label{app:modl}

Tables~\ref{tab:ct_modl} and~\ref{tab:mri_modl} give the per-grid MoDL
results summarized in Table~\ref{tab:schemes}.

\begin{table*}[t]
\centering
\caption{Sparse-view CT with MoDL under single-resolution training on
$128^2$. Entries are PSNR (dB) / SSIM ($\times100$), mean $\pm$ std over
the test set; the average is over the three grids. Best in \textbf{bold},
second best \underline{underlined}.}
\label{tab:ct_modl}
\small
\setlength{\tabcolsep}{7pt}
\resizebox{\textwidth}{!}{%
\begin{tabular}{lcccc}
\toprule
Method & $128^2$ (training grid) & $256^2$ & $512^2$ & Average \\
\midrule
FBP
& $22.66 \pm 1.86$ / $48.09 \pm 6.26$
& $23.09 \pm 1.97$ / $41.13 \pm 6.03$
& $22.98 \pm 2.05$ / $40.80 \pm 7.40$
& $22.91$ / $43.34$ \\
CNN
& $31.97 \pm 1.98$ / $84.81 \pm 3.15$
& $\mathbf{30.60 \pm 2.09}$ / $77.16 \pm 4.49$
& $26.46 \pm 1.80$ / $49.86 \pm 5.43$
& $29.68$ / $70.61$ \\
\midrule
FNO
& $28.33 \pm 1.62$ / $80.41 \pm 2.76$
& $27.19 \pm 1.61$ / $74.54 \pm 4.23$
& $26.36 \pm 1.72$ / $71.00 \pm 8.31$
& $27.29$ / $75.32$ \\
AFNO
& $32.01 \pm 1.74$ / $90.61 \pm 2.13$
& $\underline{30.11 \pm 1.86}$ / $\mathbf{83.84 \pm 4.49}$
& $\underline{27.89 \pm 2.01}$ / $\mathbf{74.38 \pm 8.81}$
& $\underline{30.00}$ / $\mathbf{82.94}$ \\
CNO
& $\mathbf{34.27 \pm 2.04}$ / $\mathbf{93.30 \pm 1.76}$
& $28.83 \pm 1.81$ / $74.53 \pm 6.96$
& $25.53 \pm 1.50$ / $50.14 \pm 6.89$
& $29.54$ / $72.66$ \\
NNO
& $\underline{32.91 \pm 1.95}$ / $\underline{91.63 \pm 2.08}$
& $29.57 \pm 1.79$ / $\underline{82.99 \pm 4.11}$
& $\mathbf{27.96 \pm 2.17}$ / $\underline{73.33 \pm 9.09}$
& $\mathbf{30.15}$ / $\underline{82.65}$ \\
\bottomrule
\end{tabular}}
\end{table*}

\begin{table*}[t]
\centering
\caption{Accelerated MRI with MoDL under single-resolution training on
$128^2$. Entries are PSNR (dB) / SSIM ($\times100$), mean $\pm$ std over
the test set; the average is over the three grids. Best in \textbf{bold},
second best \underline{underlined}.}
\label{tab:mri_modl}
\small
\setlength{\tabcolsep}{7pt}
\resizebox{\textwidth}{!}{%
\begin{tabular}{lcccc}
\toprule
Method & $128^2$ (training grid) & $256^2$ & $512^2$ & Average \\
\midrule
Zero-filled
& $30.89 \pm 2.72$ / $87.24 \pm 3.78$
& $29.46 \pm 2.75$ / $80.49 \pm 5.67$
& $28.85 \pm 2.78$ / $76.32 \pm 6.88$
& $29.73$ / $81.35$ \\
CNN
& $35.85 \pm 3.15$ / $94.04 \pm 2.64$
& $32.85 \pm 3.09$ / $87.25 \pm 5.37$
& $29.73 \pm 2.81$ / $77.41 \pm 6.58$
& $32.81$ / $86.23$ \\
\midrule
FNO
& $34.51 \pm 3.24$ / $93.68 \pm 3.35$
& $32.72 \pm 3.23$ / $88.53 \pm 5.51$
& $31.66 \pm 3.18$ / $83.92 \pm 6.88$
& $32.96$ / $88.71$ \\
AFNO
& $35.44 \pm 2.90$ / $94.73 \pm 2.00$
& $\underline{33.64 \pm 2.90}$ / $\underline{90.17 \pm 3.65}$
& $\mathbf{32.26 \pm 2.86}$ / $\mathbf{84.89 \pm 5.33}$
& $33.78$ / $\underline{89.93}$ \\
CNO
& $\underline{37.56 \pm 3.29}$ / $\underline{96.08 \pm 1.99}$
& $32.16 \pm 2.99$ / $81.75 \pm 5.71$
& $31.80 \pm 3.00$ / $81.65 \pm 6.39$
& $\underline{33.84}$ / $86.49$ \\
NNO
& $\mathbf{37.58 \pm 3.48}$ / $\mathbf{96.50 \pm 2.12}$
& $\mathbf{33.86 \pm 3.06}$ / $\mathbf{90.18 \pm 4.17}$
& $\underline{31.98 \pm 2.91}$ / $\underline{84.41 \pm 5.53}$
& $\mathbf{34.47}$ / $\mathbf{90.36}$ \\
\bottomrule
\end{tabular}}
\end{table*}

\subsection{Inference Time and Memory}
\label{app:infer_cost}

Table~\ref{tab:infer_cost} reports the time and the peak GPU memory of one
reconstruction. For CT, the time on $128^2$ is nearly the same for all
priors (235--240~ms), which indicates that it is dominated by the shared
data-consistency steps; on $512^2$, NNO takes 576~ms, compared with
474--497~ms for CNN, FNO and CNO and 704~ms for AFNO. For MRI on $512^2$,
NNO takes 96~ms, more than CNN (14~ms), FNO (37~ms) and CNO (32~ms), and
less than AFNO (199~ms). On $512^2$, NNO uses 469~MiB (CT) and 396~MiB
(MRI) of GPU memory, less than AFNO and more than the other priors.

\begin{table}[t]
\centering
\caption{Inference time (ms) / peak GPU memory (MiB) of one reconstruction
with ISTA-Net+ under single-resolution training on $128^2$, including the
data-consistency steps. One NVIDIA A100 GPU (40~GB), FP32, batch size 1.}
\label{tab:infer_cost}
\small
\setlength{\tabcolsep}{4pt}
\begin{tabular}{lccc}
\toprule
Prior & $128^2$ & $256^2$ & $512^2$ \\
\midrule
\multicolumn{4}{l}{\emph{Sparse-view CT}} \\
CNN & 238.3 / 18 & 293.3 / 59 & 473.8 / 232 \\
FNO & 239.3 / 26 & 309.5 / 42 & 489.9 / 75 \\
AFNO & 239.9 / 40 & 341.1 / 131 & 703.5 / 495 \\
CNO & 234.9 / 49 & 304.5 / 50 & 496.8 / 77 \\
NNO & 239.9 / 42 & 328.4 / 127 & 575.9 / 469 \\
\midrule
\multicolumn{4}{l}{\emph{Accelerated MRI}} \\
CNN & 20.0 / 30 & 19.9 / 59 & 13.8 / 178 \\
FNO & 31.8 / 31 & 32.0 / 54 & 37.3 / 155 \\
AFNO & 42.2 / 56 & 50.0 / 165 & 199.3 / 599 \\
CNO & 47.1 / 57 & 43.1 / 58 & 32.3 / 60 \\
NNO & 21.4 / 44 & 27.3 / 114 & 96.4 / 396 \\
\bottomrule
\end{tabular}
\end{table}

\subsection{Data Scaling}
\label{app:scaling}
\begin{figure*}[t]
\centering
\includegraphics[width=\textwidth]{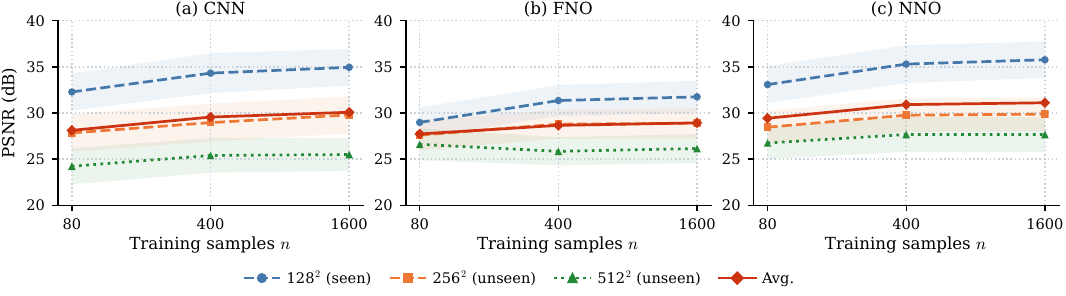}
\caption{CT data scaling within ISTA-Net+ under single-resolution training
on $128^2$. PSNR (dB); shaded bands show $\pm$ one standard deviation over
the 40 test slices.}
\label{fig:scaling}
\end{figure*}

Figure~\ref{fig:scaling} shows CT results within ISTA-Net+ when CNN, FNO and
NNO are trained on $128^2$ with $n\in\{80,400,1600\}$ slices from a larger
TCGA-LUAD set that shares no patient with the test slices; the $n=80$
subset differs from the training set of
Table~\ref{tab:ct_single}. NNO has the highest average PSNR for every $n$.
More data improves all priors on $128^2$, but the CNN prior still loses
more than 8~dB from $128^2$ to $512^2$ for every $n$.

\subsection{Sensitivity to Model Size}
\label{app:capacity}
\begin{figure*}[t]
\centering
\includegraphics[width=\textwidth]{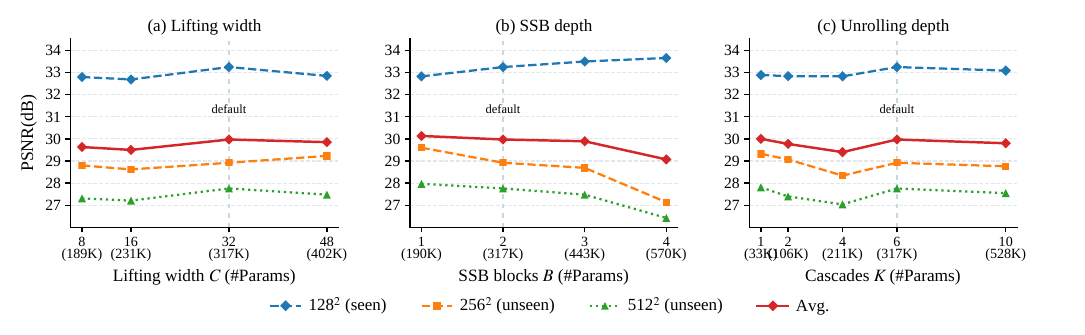}
\caption{Sensitivity of NNO to model size on sparse-view CT within
ISTA-Net+, single-resolution training on $128^2$. PSNR (dB); $C$ and $B$
in the figure denote $c$ and $L$, and parameter counts are in parentheses.}
\label{fig:capacity}
\end{figure*}

Figure~\ref{fig:capacity} varies the channel number $c$, the number $L$ of
blocks per stage and the number $K$ of stages of NNO within ISTA-Net+ on
sparse-view CT, trained on $128^2$. The average PSNR over the three grids
changes by less than 0.7~dB with $c$ and $K$, and the default ($c=32$,
$L=2$, $K=6$) is within 0.2~dB of the best average in each sweep. More
blocks per stage raise the PSNR on $128^2$ but lower it on $256^2$ and
$512^2$.

\end{document}